\documentclass[singlecolumn]{reinpaper}
\usepackage{graphicx}
\usepackage{subcaption}
\usepackage{amsmath}
\usepackage{xcolor}
\newif\ifrevhighlight\revhighlightfalse
\newcommand{\rev}[1]{\ifrevhighlight{\color{blue}#1}\else#1\fi}

\newenvironment{revblock}{\ifrevhighlight\color{blue}\fi}{}
\usepackage{algorithm}

\usepackage{algpseudocode}
\usepackage{enumitem}
\usepackage{multirow}
\usepackage{soul}
\usepackage{comment}
\usepackage{pifont}
\usepackage{caption}
\usepackage{textcomp}
\usepackage{colortbl}
\usepackage{framed}
\usepackage{tikz}
\usepackage[normalem]{ulem}
\usepackage{listings}
\usepackage{booktabs}
\usepackage{xspace}

\setlist[itemize]{align=parleft,left=0pt..1em}

\definecolor{codegreen}{rgb}{0,0.6,0}
\definecolor{codegray}{rgb}{0.5,0.5,0.5}
\definecolor{codepurple}{rgb}{0.58,0,0.82}
\definecolor{backcolour}{rgb}{0.95,0.95,0.92}
\definecolor{shadecolor}{rgb}{0.92,0.92,0.92}
\definecolor{darkgreen}{rgb}{0.0, 0.5, 0.0}

\lstdefinestyle{mystyle}{
    backgroundcolor=\color{backcolour},
    commentstyle=\color{codegreen},
    keywordstyle=\color{magenta},
    numberstyle=\tiny\color{codegray},
    stringstyle=\color{codepurple},
    basicstyle=\ttfamily\footnotesize,
    breakatwhitespace=false,
    breaklines=true,
    captionpos=b,
    keepspaces=true,
    numbers=left,
    numbersep=5pt,
    showspaces=false,
    showstringspaces=false,
    showtabs=false,
    tabsize=2
}
\newcommand{\best}[1]{\cellcolor{yellow!25}\textbf{#1}}

\newcommand{\Approach}[1]{\textit{Orion}}
\newcommand{\Healthscore}[1]{\texttt{URGE}}
\newcommand{\Threshold}[1]{\texttt{Thr}}

\renewcommand\arraystretch{0.8}

\colorlet{REINPrimary}{black}
\colorlet{REINForest}{black}
\colorlet{REINInk}{black}
\providecommand{\Description}[1]{}
\title{\textit{Orion}: Enabling Self-adaptive Memory Management for On-device Online Continual Learning}
\author{Zexin Li\textsuperscript{1}\reinorcid{0000-0001-8758-2151}\quad
Nikil Dutt\textsuperscript{2}\reinorcid{0000-0002-3060-8119}\quad
Cong Liu\textsuperscript{3}\reinorcid{0009-0001-6775-1269}}
\reinaffiliation{\textsuperscript{1}Nanyang Technological University, Singapore\\[2pt]
\textsuperscript{2}University of California, Irvine\quad
\textsuperscript{3}University of California, Riverside}
\reinemails{zexin.li@ntu.edu.sg; dutt@uci.edu; cong.liu@ucr.edu}
\reinshorttitle{Orion: Self-adaptive Memory Management}
\reintrack{ON-DEVICE ONLINE CONTINUAL LEARNING}
\reinstatus{RESEARCH PREPRINT}
\reinkeywords{Online continual learning; on-device learning; embedded systems; memory management; real-time systems; self-adaptive runtime}
\renewcommand{\reinlabname}{Orion}
\makeatletter
\renewcommand{\rein@headerleft}{ORION}
\renewcommand{\rein@footerleft}{ORION\ / \rein@status}
\renewcommand{\rein@brandrow}{%
  \noindent\begin{minipage}[c]{.28\textwidth}\centering
  \includegraphics[width=\linewidth,height=32pt,keepaspectratio]{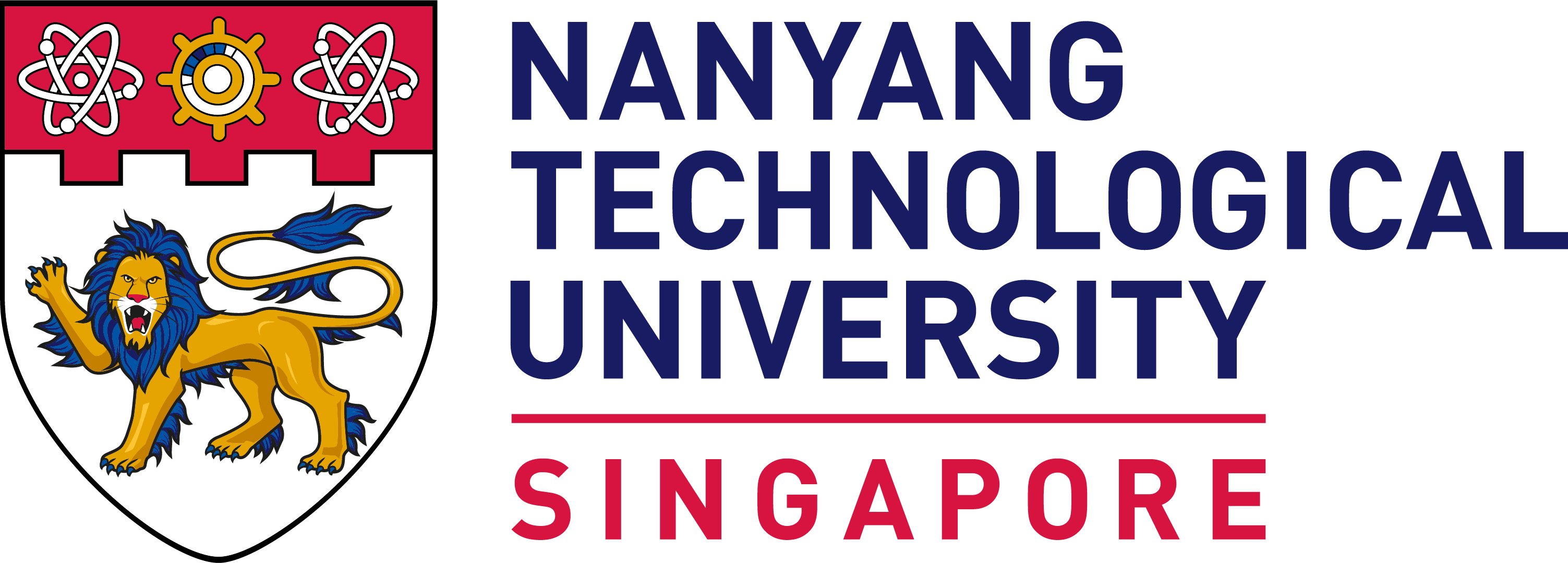}\end{minipage}\hfill
  \begin{minipage}[c]{.28\textwidth}\centering
  \includegraphics[width=\linewidth,height=32pt,keepaspectratio]{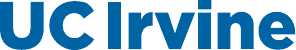}\end{minipage}\hfill
  \begin{minipage}[c]{.28\textwidth}\centering
  \includegraphics[width=\linewidth,height=32pt,keepaspectratio]{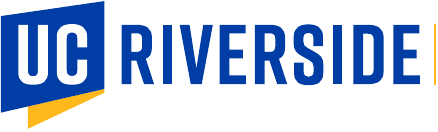}\end{minipage}\par\kern 6pt}
\makeatother

\begin{document}
\maketitle
\begin{abstract}
Online continual learning (OCL) enables real-time adaptation to new data, making it crucial for dynamic robotic applications. However, its practical deployment is hindered by memory constraints in resource-limited systems, which affect key trade-offs in training latency, plasticity, and stability. Unlike offline parameter tuning, which cannot account for the dynamic shift in memory pressure and workload complexity as OCL progresses, an online and self-adaptive approach is essential for robust on-device deployment. This paper proposes \Approach{}, a holistic framework designed to co-optimize training latency, plasticity, and stability of state-of-the-art OCL models under strict memory constraints, enabling feasible on-device deployment. At its core, \Approach{} leverages URGE, a unified runtime indicator grounded in the ``Buckets effect'' principle that system performance is bounded by its scarcest resource, to dynamically reallocate memory across OCL components by jointly coordinating batch processing, replay buffers, and optimization strategies\rev{; URGE compares every metric with an automatic set-point on a dimensionless scale, and a hard constraint keeps every allocation within the device memory budget}. A system prototype of \Approach{} has been implemented using the widely adopted \texttt{Avalanche-lib} and \rev{evaluated across five benchmarks, four OCL algorithms, three backbones, and three memory regimes (an edge server, an NVIDIA Jetson AGX Orin, and the Orin under an 8\,GB budget), with three seeds each.} To further demonstrate its practical utility, \Approach{} is integrated into a realistic autonomous navigational robot powered by OCL. \rev{\Approach{} completes every run within the memory budget, while fixed configurations run out of memory; where all methods fit, it reaches the highest final accuracy on all five benchmarks and trains 9--16$\times$ faster than the Avalanche defaults, and it is more accurate than Miro, the closest runtime system, in most settings. Its controller costs about 1\% of the training time and energy on typical streams, making \Approach{} a practical solution for on-device continual learning.}
\end{abstract}
\section{Introduction}

Online continual learning (OCL) addresses the challenge of learning from non-stationary, streaming data while mitigating catastrophic forgetting~\citep{chaudhry2019continual,prabhu2023computationally,lopez2017gradient,chaudhry2018efficient,aljundi2019gradient,kirkpatrick2017overcoming,li2017learning,soutif2023comprehensive}. This capability is critical for robotics and other embodied AI systems that must adapt in real time in dynamic environments such as transportation, agriculture, and defense~\citep{nie2023online,vodisch2023covio,vodisch2023codeps,castri2023continual,hajizada2024continual}. Unlike offline retraining on fixed datasets, OCL updates the model incrementally in a single pass over new experiences, making both algorithmic behavior and systems design central to practical deployment~\citep{mai2022online,aljundi2019online}.

OCL performance is commonly characterized by \emph{plasticity} (fast adaptation to new experiences) and \emph{stability} (retention of prior knowledge), together with \emph{training latency}, i.e., how quickly the model incorporates new data~\citep{ghunaim2023real}. On-device deployment adds a fourth, implicit axis: \emph{memory}.  Recently, replay-based methods such as ER~\citep{chaudhry2019continual} revisit stored experiences and are effective across diverse settings, including resource-constrained platforms~\citep{prabhu2023computationally}. Memory selection (GSS) favors informative samples~\citep{aljundi2019gradient}. Constraint- and regularization-based methods, e.g., GEM~\citep{lopez2017gradient}, AGEM~\citep{chaudhry2018efficient},  EWC~\citep{kirkpatrick2017overcoming}, and LwF~\citep{li2017learning}, preserve past knowledge via gradient projections or parameter consolidation. Beyond algorithms, system-level efforts remain limited: LifeLearner targets hardware-level optimization for meta-continual learning~\citep{kwon2023lifelearner}, Ekya optimizes \emph{offline} CL pipelines on multi-GPU servers~\citep{bhardwaj2022ekya}, and Latent Replay focuses on algorithmic changes without systems co-design~\citep{pellegrini2020latent}. 
Yet none of these efforts jointly address runtime memory constraints, algorithmic efficacy, and system-level co-design for \emph{on-device} OCL. This gap is critical, as autonomous robots and edge devices increasingly require real-time adaptation under strict hardware budgets in the field, far from data centers and without opportunity for offline retraining.
 
\begin{figure}[!tbp]
\centering
    \includegraphics[width=0.7\textwidth]{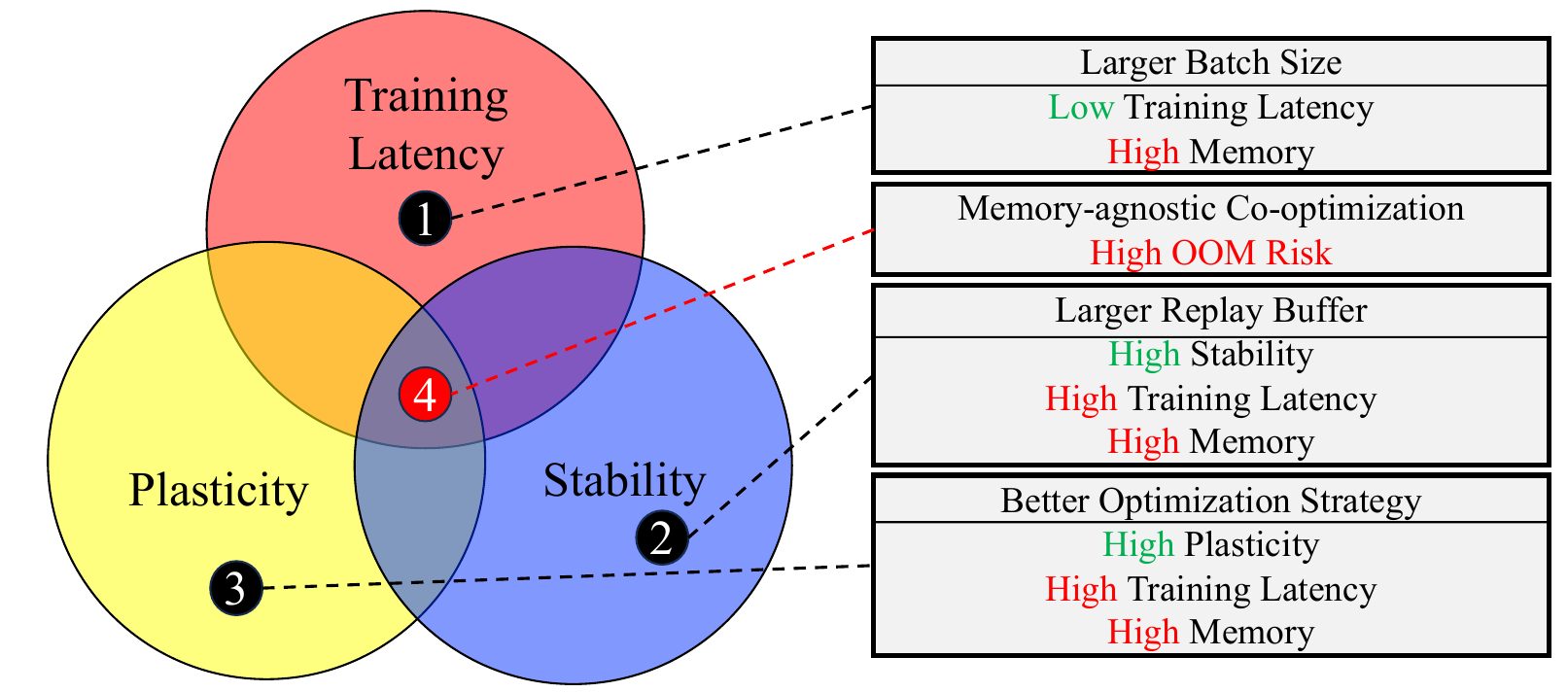}
    \caption{\textbf{No silver bullet.} Visualization of the complex tradeoff space among three key performance metrics of on-device OCL. Memory-agnostic co-optimizing training latency, plasticity, and stability could easily raise out-of-memory (OOM) concerns under stringent memory constraints.}
    \label{fig:vis-tradeoff-space}
    
\end{figure}

More challenging, the trade-off space of on-device OCL is inherently conflicted: batch scaling reduces latency but inflates memory, larger buffers help stability, but risk OOM, and stronger optimization improves plasticity while increasing compute and memory traffic. Figure~\ref{fig:vis-tradeoff-space} summarizes this “no silver bullet’’ landscape and motivates an adaptive, memory-aware runtime.
Critically, because memory pressure and workload complexity shift dynamically as the number of OCL experiences grows, any static offline configuration could be either too conservative and waste resources early, or too aggressive and cause OOM failures later. This necessitates an online, self-adaptive memory manager.

\noindent\textbf{Contributions.}
This work introduces \Approach{}, a self-adaptive memory management ecosystem for on-device OCL that addresses the challenge of balancing training latency, plasticity, and stability under stringent memory constraints. Managing these trade-offs is non-trivial due to OS-level allocation and runtime variability; doing so robustly requires tight integration of application- and system-level decisions, especially on embedded platforms.

At its core, \Approach{} introduces \Healthscore{}, 
a unified runtime indicator that \rev{steers the OCL process online, without per-deployment tuning: a new device, benchmark, or algorithm requires only the device memory budget and, optionally, an importance order of the metrics. URGE compares every metric with an automatically chosen set-point on a dimensionless scale, and is evaluated separately for each memory consumer. Its formulation is} intentionally pragmatic and interpretable, inspired by Liebig's Law of the Minimum~\citep{odum1971fundamentals}, \rev{and we evaluate its robustness across benchmarks, hardware platforms, OCL algorithms, backbones, and hyperparameter settings.} By integrating performance metrics (plasticity, stability, latency), balancing memory usage with system constraints to prevent OOM errors, and adapting to workload changes using time-dependent thresholds, \Healthscore{} enables autonomous memory reallocation across OCL components. This self-adaptive mechanism ensures efficient resource utilization, enabling \Approach{} to maintain OCL algorithmic performance while adhering to strict resource constraints. In addition, we prototype a practical system for real-world deployment by extending the \texttt{Avalanche-lib} library~\citep{avalanche1,avalanche2} with system-level optimizations tailored for embedded and edge platforms. \rev{The prototype loads streaming and replay data in parallel to training, through a single prefetching path for both.} All enhancements are implemented as transparent modules, enabling rapid deployment of \Approach{} across diverse hardware, including ARM64-based embedded devices and edge servers.

\noindent\textbf{Implementation and Evaluation.}
\Approach{} is evaluated on standard OCL benchmarks~\citep{krizhevsky2009learning,lomonaco2019fine,lomanco2017core50,hess2021procedural} and four representative algorithms~\citep{chaudhry2019continual,aljundi2019gradient,lopez2017gradient,chaudhry2018efficient}, \rev{with three backbones and three memory regimes: an edge server, an NVIDIA Jetson AGX Orin~\citep{bib:orin}, and the Orin under an emulated 8\,GB budget. Every result is the mean over three seeds, and \Approach{} is compared with the Avalanche defaults, latent replay, and Miro~\citep{ma2023cost}, the closest runtime system.} We further demonstrate practicality in an autonomous driving case study built on \texttt{EndlessCL-Sim}~\citep{hess2021procedural} with a navigational TurtleBot~3.

Key highlights of \Approach{} include:

\begin{itemize}[leftmargin=10pt]
\item \textbf{Overall Effectiveness and Versatility:} \rev{\Approach{} completes every run within the memory budget, whereas fixed configurations run out of memory as soon as they do not fit (e.g., MAX-P on every benchmark under 8\,GB, MAX-A and LR with GSS on the 32\,GB Orin). Where all methods fit, \Approach{} reaches the highest final accuracy on all five benchmarks with ER (9.4 points above the Avalanche defaults on average) and trains 16.3$\times$ (edge server) and 9.0$\times$ (Orin) faster than them~(Sec.~\ref{sec:overall_effectiveness}). It is also more accurate than Miro in most settings, at a comparable total latency~(Sec.~\ref{sec:miro_comparison}).}

\item \rev{\textbf{Adaptation without offline tuning:} The ablation isolates \Healthscore{}-driven adaptation: it improves on the same configuration kept fixed and on a memory-only controller, and without any offline run it matches a configuration tuned with nine offline runs on SplitCIFAR100 and closes two thirds of the gap to it on CORe50-NC, while offline tuning remains better on the longest stream~(Sec.~\ref{sec:ablation}); a hard constraint guarantees memory compliance, and the controller is robust to its hyperparameters~(Sec.~\ref{sec:sensitivity}).}

\item \textbf{Practical Usability:} A robotic case study validated the practicality and efficacy of \Approach{} in real-world OCL scenarios. \rev{It completes all twelve algorithm--scenario combinations on the Jetson AGX Orin and is the fastest method in eleven of them~(Sec.~\ref{sec:oclrobot_case_study}).}

\item \textbf{Low Overhead:} \rev{\Approach{}'s controller costs about 1\% of the training time and energy on typical streams (6--7\% on a stream with 79 experiences), and its memory overhead is negligible~(Sec.~\ref{sec:overhead}).}
\end{itemize}

\section{Background}
\label{sec:background}

\subsection{Online Continual Learning}
\label{sec:background_OCL}

Online Continual Learning (OCL) enables systems to incrementally learn from streaming data, adapting to new information while retaining prior knowledge~\citep{mai2022online}. Unlike traditional deep learning, which relies on static datasets and offline retraining, OCL is essential for dynamic, real-world applications where data evolves, and full dataset retraining is impractical. OCL processes sequential experiences, new batches of data, by incrementally updating model parameters and balancing adaptation to new data with retention of prior learned knowledge. In real-world robotic applications, OCL provides a lightweight mechanism for adapting to new lighting or terrains in autonomous navigation, recognizing novel objects on the fly~\citep{pellegrini2020latent}, updating person re-identification models under varying conditions~\citep{ye2024person}, and continuously refining physical manipulation skills~\citep{lee2024incremental}.

OCL introduces unique challenges due to its online workload characteristics. It cannot rely on multiple runs or offline optimization methods, thus inherently requiring \textit{fast, adaptive system-application coordination}. Unlike offline methods, OCL demands rapid, incremental learning, which makes it particularly challenging for resource-constrained devices where low training latency is critical~\citep{aljundi2019online}. 
Following state-of-the-art OCL algorithms~\citep{ghunaim2023real,prabhu2023computationally}, we evaluate on-device OCL performance using the following metrics
\begin{itemize}[leftmargin=10pt]
\item Plasticity: the ability to adapt to new data.
\item Stability: the ability to retain accuracy on previously learned data.
\item  Training Latency: The speed of adaptation to new data, where higher latency may indicate slower adaption and reduced accuracy~\citep{ghunaim2023real}.
\item Memory: adherence to stringent memory constraints, which is crucial for on-device deployment. 
\end{itemize}

Formally, given a dataset $\mathcal{D}$ and experience number $\mathcal{N}$, we define  plasticity (P), and stability (S) for the $\mathcal{K}$-th experience $(\mathcal{K} \leq \mathcal{N})$ as follows: Plasticity (P) indicates the ability to adapt to new knowledge: 
$ P = \frac{1}{\mathcal{K}} \sum_{i=1}^{\mathcal{K}} \text{acc}_i$,
where \text{acc}${_i}$ \text{ is the test accuracy on the } $i$-th experience. Stability (S) indicates the ability to maintain previously learned knowledge while integrating new knowledge:
\begin{align*}
    S &= 
    \begin{cases} 
        1 & \text{if } \mathcal{K} = 1, \\
        1 - \frac{1}{\mathcal{K}-1} \sum_{i=1}^{\mathcal{K}-1} \mathcal{F}_i & \text{if } \mathcal{K} > 1.
    \end{cases}
\end{align*}
where $\mathcal{F}_i$ measures how much knowledge about previous experiences is lost after learning the $i$-th experience. 
This equation also considers that when \( \mathcal{K} = 1 \), there are no prior experiences to evaluate forgetting, so stability is set to one. 
Plasticity and stability can be assessed immediately after training in each experience, without needing to complete the entire OCL process.

To gain a comprehensive understanding of the performance of on-device OCL, it is essential to focus on two key OCL components: batch execution and replay buffer. Batch execution involves processing multiple data samples in a single operation, fostering effective learning. On the other hand, the replay buffer serves as a repository for past experiences, which are sampled for conducting mini-batch training. These components significantly impact algorithmic performance, that is, plasticity and stability, and system-level memory usage and training latency. Specifically, we explore the impacts of key OCL parameters on the above metrics, including (1) training batch size, which defines the amount of data used in  each mini-batch training; 
(2) replay buffer size, which determines the amount of data stored for revisitation; and (3) optimization strategies tailored for enhancing plasticity.

\section{Motivation}
\label{sec:motivation}

This section thoroughly examines the critical trade-offs in on-device OCL under stringent memory constraints, highlighting specific challenges through a series of realistic case studies. Understanding and quantifying these trade-offs provides valuable insights into how to effectively co-optimize plasticity, stability, and resource usage, guiding the development of more robust and efficient deployable solutions.

On-device OCL must jointly balance \emph{training latency}, \emph{plasticity}, and \emph{stability} under tight memory budgets. We study three levers that practitioners frequently adjust in practice (batch size, replay buffer size, and optimization plugins), and find that memory is the shared bottleneck that frequently drives failures on embedded platforms.

\begin{figure*}[!tbp]
    \centering
    \begin{subfigure}[t]{0.48\textwidth}
        \centering
        \includegraphics[width=\linewidth]{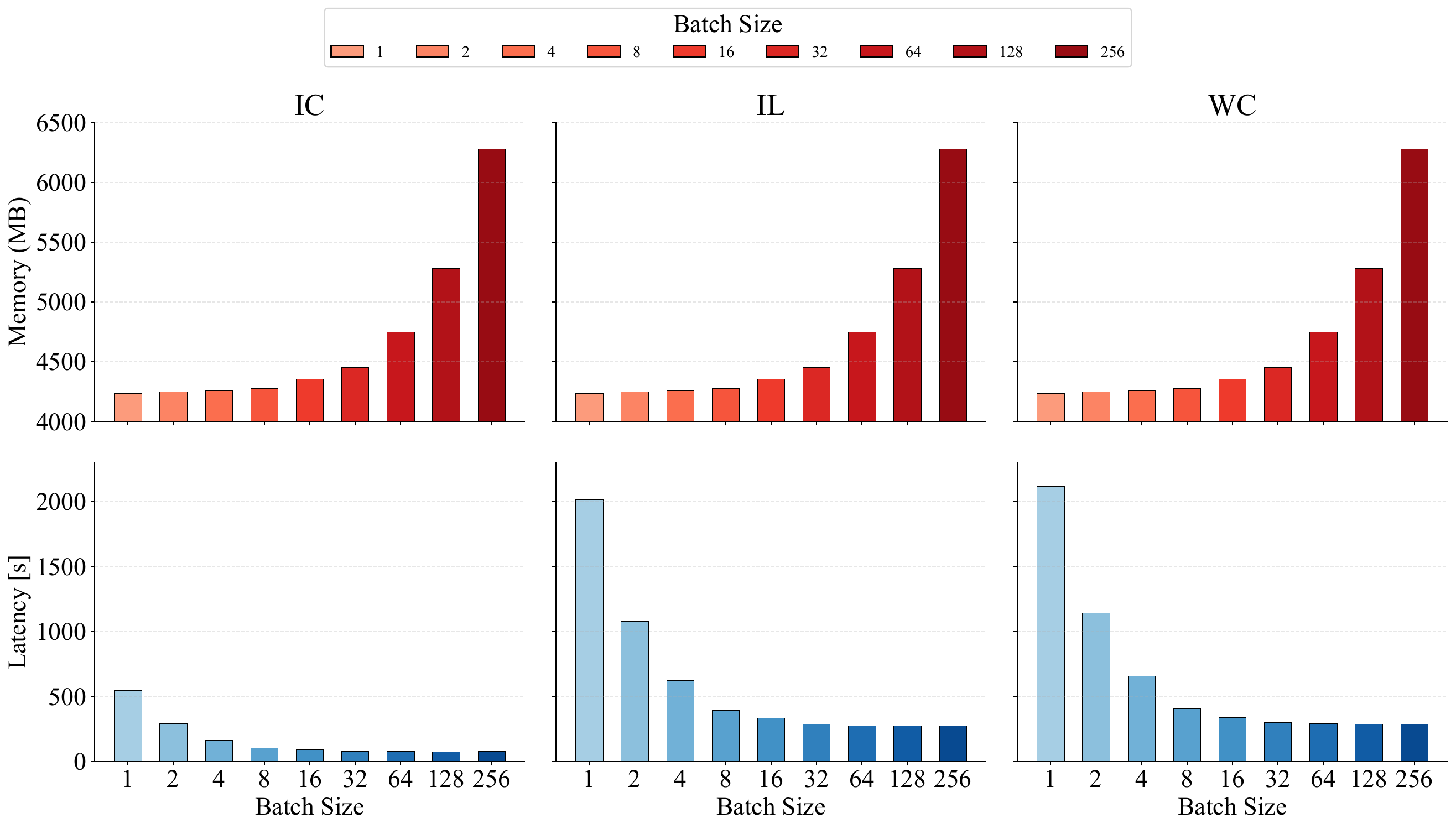}
        \caption{Effect of training batch size on OCL metrics.}
        \label{fig:batchsize}
    \end{subfigure}
    \hfill
    \begin{subfigure}[t]{0.48\textwidth}
        \centering
        \includegraphics[width=\linewidth]{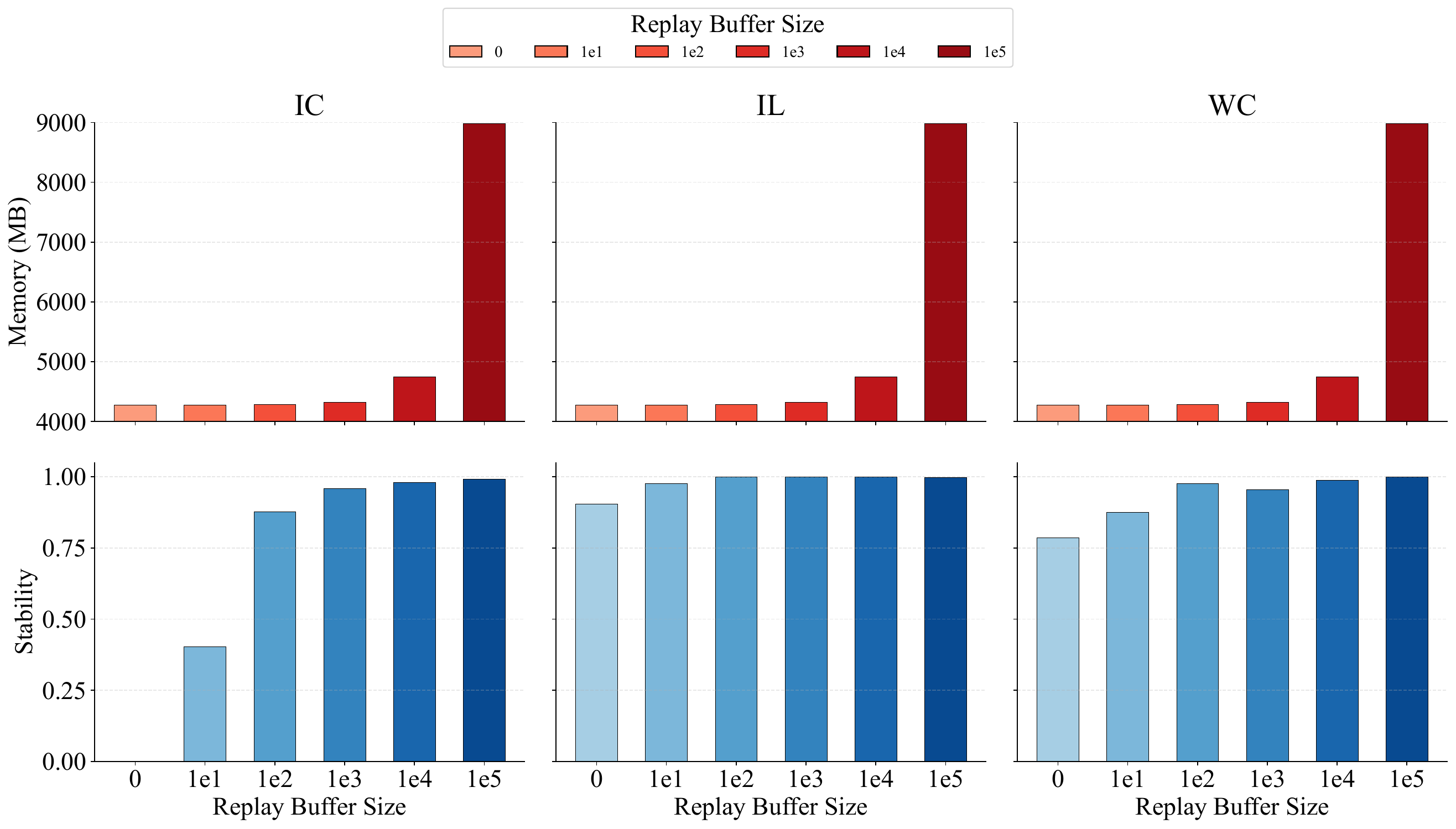}
        \caption{Effect of replay buffer size on OCL metrics.}
        \label{fig:replaybuffersize}
    \end{subfigure}
    
    \caption{Impact of key hyperparameter choices on OCL metrics under memory constraints.}
    \label{fig:combined_hyperparameters}
    
\end{figure*}

\begin{table*}[!tbp]
    \centering
    \renewcommand\arraystretch{0.9}
    \caption{Integrating OCL plugins into ER. Best results are marked. Subscripts show differences from the baseline.}
    \resizebox{\textwidth}{!}{
    \begin{tabular}{l|cccc|cccc|cccc}
        \toprule
        Metrics & \multicolumn{4}{c|}{Training Latency (s)} & \multicolumn{4}{c|}{Plasticity} & \multicolumn{4}{c}{Memory (MB)} \\
        \midrule
        Setting & w/o Opt. & GEM & EWC & GEM+EWC & w/o Opt. & GEM & EWC & GEM+EWC  & w/o Opt. & GEM & EWC & GEM+EWC \\
        \midrule
        IC & \best{73.06} & 147.40$_{+74.34}$ & 147.18$_{+74.12}$ & 215.13$_{+142.07}$ & 0.98 & \best{0.99}$_{+0.01}$ & \best{0.99}$_{+0.01}$ & \best{0.99}$_{+0.01}$ & \best{4100} & 4195$_{+95}$ & 4200$_{+100}$ & 4207$_{+107}$ \\
        IL & \best{263.28} & 505.00$_{+241.72}$ & 537.28$_{+274.00}$ & 771.61$_{+508.33}$ & 0.94 & 0.94$_{+0.00}$ & \best{0.98}$_{+0.04}$ & 0.93$_{-0.01}$ & \best{4250} & 4347$_{+97}$ & 4352$_{+102}$ & 4360$_{+110}$ \\
        WC & \best{268.88} & 527.16$_{+258.28}$ & 560.63$_{+291.75}$ & 817.82$_{+548.94}$ & 0.97 & 0.89$_{-0.08}$ & 0.93$_{-0.04}$ & \best{0.97}$_{+0.00}$ & \best{4180} & 4270$_{+90}$ & 4281$_{+101}$ & 4301$_{+121}$ \\
        \bottomrule
    \end{tabular}}
    
    \label{tab:optimized_ocl}
\end{table*}

\rev{\noindent\textbf{Three levers, one memory pool.} We vary the three levers one at a time on the Endless-Sim scenarios (Fig.~\ref{fig:combined_hyperparameters}, Table~\ref{tab:optimized_ocl}). \emph{Batch size} trades latency for memory: batches of 1--16 keep the footprint near 4.2\,GB but take over 2000\,s in IL and WC, while a batch of 256 cuts latency below 200\,s and raises the footprint beyond 6\,GB. \emph{Replay capacity} has diminishing returns: stability saturates near 1.0 beyond $10^4$ samples, while the footprint jumps to about 9\,GB at $10^5$. \emph{Optimization plugins} (GEM, EWC) slightly improve plasticity but double or triple latency and add 90--121\,MB.}

\begin{figure}[!tbp]
    \centering
    \includegraphics[width=\textwidth]{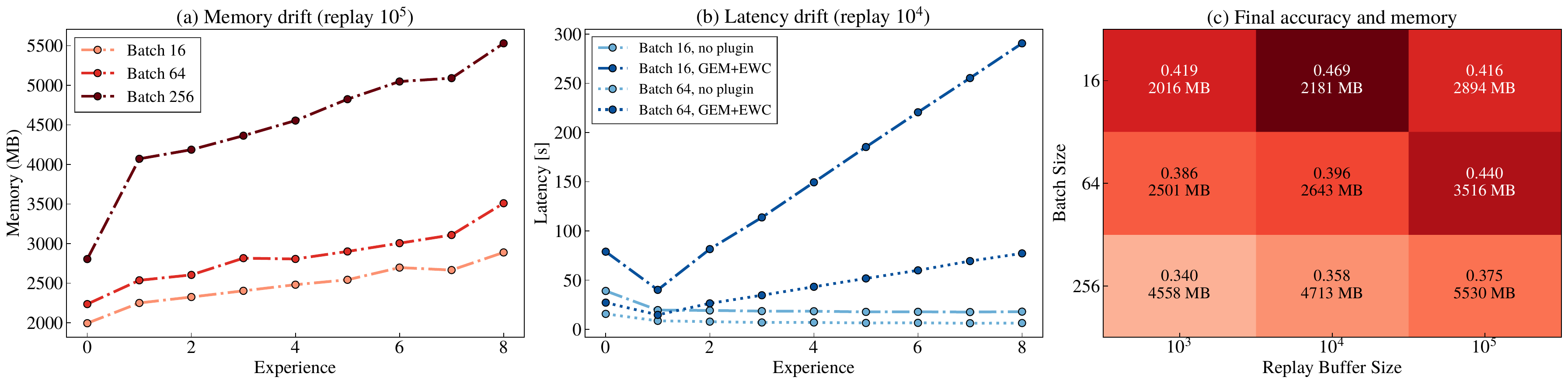}
    \caption{\rev{Joint effect of the levers and temporal drift (CORe50-NC, ER, Jetson AGX Orin, fixed configurations). (a) Peak footprint per experience with $R=10^5$; (b) training latency per experience with and without the GEM+EWC plugins; (c) final average accuracy (top) and peak footprint (bottom) of the $B \times R$ grid without plugins.}}
    \label{fig:drift}
\end{figure}

\begin{revblock}
\noindent\textbf{The levers interact, and their footprint drifts.} The single-lever view understates the problem, because the levers compete for the same memory and their costs change along the stream. Fig.~\ref{fig:drift} runs 18 fixed configurations (batch size $B \in \{16, 64, 256\}$, replay capacity $R \in \{10^3, 10^4, 10^5\}$, plugins on or off) on CORe50-NC with ER on the Jetson AGX Orin and records the footprint and latency of every experience. Three observations follow. (a) \emph{The footprint drifts along the stream.} With $R=10^5$, the peak footprint grows by 45\% ($B=16$), 57\% ($B=64$), and 97\% ($B=256$) between the first and the last experience, as the buffer fills and the replay data loader allocates more memory; a configuration sized on the first experience can exceed the budget later. (b) \emph{The latency drifts too, depending on the plugins.} Without plugins, the per-experience latency is flat after the first experience; with GEM and EWC, it grows linearly with the number of past experiences, from 40\,s to 291\,s per experience at $B=16$ and $R=10^4$, because both plugins keep per-experience state that every training step must process. (c) \emph{The levers interact, and the best combination is not the corner of either.} The joint footprint grows with both $B$ and $R$ (from 2.0\,GB to 5.5\,GB across the grid), but accuracy is not monotone in either lever: the most accurate configuration is $B=16$, $R=10^4$ (0.469), a larger buffer helps at $B=64$ and $B=256$ but hurts beyond $10^4$ at $B=16$, and enabling the plugins lowers the final accuracy in every one of the nine combinations (by 7.2 to 17.8 points) while multiplying the latency by 1.8--8.4$\times$. A configuration that fits and is fast at the beginning of the stream is therefore not guaranteed to fit, or to be fast, at its end, and the best configuration cannot be read off a single experience. This is the premise for an online controller rather than a configuration chosen once.
\end{revblock}

\noindent\textbf{Takeaway:} Taken together, these three patterns all draw from the same memory pool. When batch size increases, when the buffer grows, or when heavier plugins are enabled, the combined footprint often exceeds device capacity, which leads to OOM and unstable training. Managing these trade-offs is a systems problem. Operating system memory allocation interacts with algorithm choices; static or memory blind tuning is brittle. We need a fast and self-adaptive memory manager that coordinates batch processing, replay buffers, and plugin usage to avoid OOM while preserving stability and training efficiency. The next section develops this design.
\section{System Design}

\subsection{Design Overview}

Our proposed system, named \Approach{}, is a holistic framework designed to optimize OCL systems through self-adaptive runtime memory management, balancing system performance and resource constraints. Importantly, \Approach{} is transparent to the online continual learning process and can be seamlessly integrated into existing OCL frameworks. This ensures that training and evaluation steps remain consistent with established practices in the field, aligning with the methodologies used in state-of-the-art continual learning studies~\citep{aljundi2019gradient,chaudhry2018efficient}. Specifically, \Approach{} maintains transparency between data and experience splits in benchmark evaluations, adhering to the protocols followed in prior continual learning works.

\begin{figure*}[!tbp]
\centering
    \includegraphics[width=0.85\textwidth]{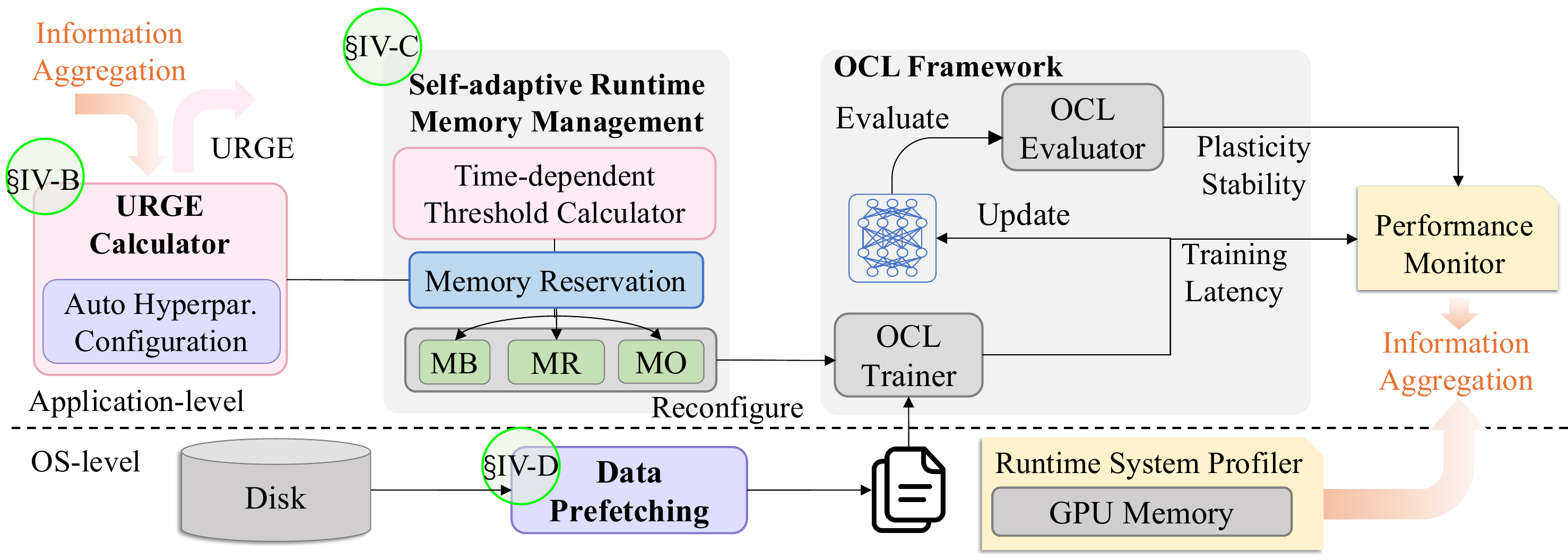}
    \caption{Design overview of \Approach{}.}
    \label{fig:design_overview}
    
\end{figure*}

As shown in Fig.~\ref{fig:design_overview}, \Approach{} employs a hierarchical control framework guided by a unified, system-aware indicator (\Healthscore{}). This indicator integrates multiple performance metrics to provide a comprehensive evaluation of the system’s state and serves as the foundation for decision-making. Intuitively, \Healthscore{} captures the trade-offs between system performance and memory resource availability. A high \Healthscore{} value indicates that resources are abundant but the system is underperforming, suggesting both the opportunity and urgency for optimization without significant risk of overloading memory. Conversely, a low \Healthscore{} value reflects either: (1) the system is performing sufficiently well and does not require further optimization, or (2) resource constraints that make further optimization impractical or risky due to potential OOM errors.
(Detailed calculation and application of \Healthscore{} is in Sec.~\ref{sec:health_score}.) \rev{\Healthscore{} expresses a \emph{preference} for growing or shrinking the allocations; feasibility is not left to \Healthscore{}: every allocation is projected onto the memory budget by a hard constraint (Sec.~\ref{sec:fine_grained_control}).}

\Approach{} implements a hierarchical control framework which operates at two levels:
(1) Coarse-grained control (\rev{Algorithm~\ref{alg:healthscore_memory}, lines 4--6}): At this level, \Healthscore{} is calculated to guide high-level decisions about the overall \rev{optimization direction} of the OCL system. 
(2) Fine-grained control (\rev{Algorithm~\ref{alg:healthscore_memory}, lines 7--12}): This level \rev{translates the decision into memory allocations for batch processing, the replay buffer, and the optimization plugin, and projects them onto the device memory budget.}
\rev{At each experience boundary, \Healthscore{} is computed from the measurements of the experience just completed} and incorporated with a time-dependent threshold to make adaptive memory management decisions, as detailed in Section~\ref{sec:fine_grained_control}.  

Beyond its adaptive control logic, \Approach{} \rev{is implemented as an engineering prototype that loads stream and replay data in parallel to training through a single prefetching path.} All system-level optimizations are implemented as transparent, drop-in modules within our prototype, ensuring compatibility with the standard Avalanche pipeline and minimal engineering overhead for end users. Details of these engineering contributions are described in Section~\ref{sec:prototype_implementation}. 

\subsection{The Design of \Healthscore{}}
\label{sec:health_score}

\rev{The design of \Healthscore{} is inspired by Liebig's Law of the Minimum (also known as the ``Buckets effect'')~\citep{odum1971fundamentals}, which posits that growth is limited not by the total resources available but by the scarcest resource (limiting factor). In this context, \Healthscore{} ensures that the system achieves a balance among plasticity, stability, and latency while respecting the memory budget. When plasticity and stability are at the best level the learner has reached, when latency is at its reference level, or when memory headroom is scarce, \Healthscore{} decreases, leading to more conservative memory allocation. Conversely, when the learner falls behind on a metric and memory headroom is available, \Healthscore{} increases, allowing more aggressive memory allocation.}

\begin{revblock}
\noindent\textbf{Dimensionless terms.} Let $P_t, S_t \in [0,1]$ be the plasticity and stability measured after experience $t$, $\ell_t$ the training latency per processed sample (seconds), and $M_t$ the peak memory footprint during experience $t$ (MB), with device budget $M_{max}$ and headroom $h_t = 1 - M_t / M_{max}$. These quantities live on different scales, so instead of comparing raw values with thresholds, \Approach{} first converts every metric into a \emph{relative} deviation from its set-point:
\begin{equation}
d^P_t = \frac{P^{ref}_t - P_t}{P^{ref}_t}, \quad
d^S_t = \frac{S^{ref}_t - S_t}{S^{ref}_t}, \quad
d^L_t = \frac{\ell_t - \ell^{ref}}{\ell^{ref}}, \quad
d^M_t = \frac{h_t - h_{safe}}{h_{safe}} .
\label{eq:deviations}
\end{equation}
All four deviations are dimensionless: $d^P_t = 0.1$ means plasticity is 10\% below its set-point, $d^L_t = 0.1$ means latency is 10\% above its reference, independently of whether latency is measured in seconds or milliseconds and memory in MB or GB. Each deviation is mapped to $(0,1)$ by a sigmoid $u^x_t = \sigma(\kappa\, d^x_t)$, $\sigma(z) = 1/(1+e^{-z})$, except memory, which can only restrain: $u^M_t = \sigma(\kappa \min(0, d^M_t))$ equals $0.5$ (neutral) while the headroom exceeds the safety margin and decreases towards $0$ as the headroom is exhausted, so abundant memory alone never triggers growth. \Healthscore{} is the weighted geometric mean of the terms:
\begin{equation}
\Healthscore{}_t = \prod_{x \in \{P,S,L,M\}} \left(u^x_t\right)^{w_x}
= \prod_{x} \sigma\!\left(\kappa\, d^x_t\right)^{w_x}, \qquad \sum_x w_x = 1 .
\label{eq:health_score}
\end{equation}
The terms keep their original meaning: $u^P_t$ ($u^S_t$) is high when plasticity (stability) has fallen below its set-point, $u^L_t$ is high when training has become slower, and $u^M_t$ drops below $0.5$ once the headroom falls under the safety margin $h_{safe}$. A single gain $\kappa$ applies to all terms, because the arguments are commensurable, and the user's importance order enters only through the exponents $w_x$ (Sec.~\ref{sec:auto_hyper}). Using exponents rather than slopes has two consequences that the original product of raw-unit sigmoids did not have. First, a weight scales the influence of a term without changing how sharply the term reacts, so the importance order translates into the intended behavior regardless of units. Second, \Healthscore{} has a fixed neutral point: when every metric is at its set-point, all terms equal $0.5$ and so does \Healthscore{}, whatever the weights, which gives the threshold below a scale-free meaning.

\noindent\textbf{Set-points.} \Approach{} does not require dataset-specific targets. The plasticity and stability set-points are the best values the learner has reached so far in the stream, $P^{ref}_t = \max_{\tau \le t} P_\tau$ and $S^{ref}_t = \max_{\tau \le t} S_\tau$, so the terms measure regression from what the current configuration history has already demonstrated. The latency reference $\ell^{ref}$ is the per-sample latency of the first experience under the initial configuration, or a latency target if the application specifies one. The headroom set-point $h_{safe}$ is the safety margin of the memory budget, which is also used by the hard constraint in Sec.~\ref{sec:fine_grained_control}. Hence the only quantity the user must provide is the memory budget $M_{max}$ of the device.

\noindent\textbf{Relation to the original formulation.} The original \Healthscore{} applied the sensitivities directly to raw differences such as $k_l (L_t - L_{th})$ and $k_m (M_t - M_{max})$. Writing $k_l (L_t - L_{th}) = (k_l L_{th})\, d^L_t$ shows that the effective sensitivity of each raw-unit term is its gain multiplied by the scale of its set-point, which is exactly the unit dependence that the normalization above removes.

\noindent\textbf{One indicator per memory consumer.} The three knobs do not move the metrics in the same direction (Sec.~\ref{sec:motivation}): a larger batch lowers latency but also plasticity, because a single pass then performs fewer updates; a larger replay buffer raises stability at a memory cost; the optimization plugins can raise plasticity and stability at a latency cost. Driving all knobs with the same scalar, as in the submitted version, therefore grows the batch exactly when plasticity is deficient. \Approach{} instead evaluates Eq.~\ref{eq:health_score} once per memory consumer $k \in \{B, R, O\}$ over the terms that consumer affects, using $u^x_t$ when the knob relieves deficit $x$ and $1-u^x_t$ when it aggravates it:
\begin{equation}
\begin{aligned}
\Healthscore{}^{B}_t &= \left(u^L_t\right)^{\tilde w_L} \left(1-u^P_t\right)^{\tilde w_P} \left(u^M_t\right)^{\tilde w_M}, \\
\Healthscore{}^{R}_t &= \left(u^S_t\right)^{\tilde w_S} \left(u^M_t\right)^{\tilde w_M}, \\
\Healthscore{}^{O}_t &= \left(u^P_t\right)^{\tilde w_P} \left(u^S_t\right)^{\tilde w_S} \left(1-u^L_t\right)^{\tilde w_L} \left(u^M_t\right)^{\tilde w_M},
\label{eq:knob_urge}
\end{aligned}
\end{equation}
where $\tilde w$ are the user weights renormalized over the terms of each consumer. All three indicators share the neutral point $0.5$, the threshold, and the hard constraint below; the scalar \Healthscore{}$_t$ of Eq.~\ref{eq:health_score} is retained as an ablation (Sec.~\ref{sec:ablation}).
\end{revblock}

\subsection{Self-adaptive Memory Management}
\label{sec:fine_grained_control}

This section explains how \Approach{} leverages \Healthscore{} for self-adaptive memory management. By dynamically adjusting memory allocation in response to performance metrics, resource constraints, and workload complexity, \Approach{} ensures efficient and robust optimization for on-device OCL systems.

\noindent \textbf{Time-dependent Threshold.}
While performance metrics such as stability and plasticity are only accessible after the completion of training for one experience rather than during training, they remain instrumental for guiding fine-grained control. This is particularly relevant in OCL, where workloads dynamically evolve. As the OCL progresses, each experience typically exhibits increasing training latency due to the growing computational complexity required to retain old knowledge (\rev{Fig.~\ref{fig:drift}}). This dynamic is distinct from the relatively homogeneous patterns in training latency observed during each training epoch of the standard Deep Neural Network fine-tuning process~\citep{goyal2017accurate}. To account for this, we introduce a time-dependent threshold, $\Threshold{}_{t}$, that decreases over time to encourage a more aggressive optimization policy as training advances:
\begin{equation}
\rev{\Threshold{}_{t} = \Threshold{}_{min} + \left(\Threshold{}_{0} - \Threshold{}_{min}\right) e^{-\delta (t - t_0)}}
\label{eq:threshold}
\end{equation}
\rev{where $\Threshold{}_{0}$ is the initial threshold value and $\delta$ is the decay rate.} \rev{Because \Healthscore{} is neutral at $0.5$, we start slightly above it ($\Threshold{}_0 = 0.6$, conservative while the set-points are still being established) and decay towards a floor $\Threshold{}_{min} = 0.4$. The floor keeps the controller able to shrink allocations late in the stream; with the original exponential decay to zero, the controller would eventually grow allocations whatever the state. $t_0$ is the start of the current schedule ($t_0=0$, or the last detected distribution shift, see below).}

\rev{\noindent\textbf{Unified \Healthscore{}-based Memory Management.} We propose a unified memory manager for application-aware memory allocation and reallocation. The memory manager uses \Healthscore{} and system memory availability to make decisions about memory reservations for each OCL component and reactively adjusts application-level control knobs (batch size, replay buffer size, and optimization strategies).
\rev{All updates happen at the boundary before experience $t \geq 1$ is trained, using the measurements of experience $t-1$; experience $0$ is trained with the initial allocation $(MB_0, MR_0, MO_0)$ because nothing has been measured yet.}
The memory allocated for batch processing is adjusted based on \Healthscore{}:}
\begin{equation}
MB_t = \begin{cases}
MB_0, & \text{if } t = 0 \\
MB_{t-1} \cdot \rev{\mathrm{clip}_{[1/s,\, s]}}\!\left(1 + \alpha \cdot (\rev{\Healthscore{}^{B}_t} - \Threshold{}_{t})\right), & \text{if } t > 0
\end{cases}
\label{eq:mb}
\end{equation}
\rev{where $\alpha$ controls the sensitivity of the batch processing memory to \Healthscore{}}\rev{, and the step is limited to a factor $s=2$ per experience so that a single noisy measurement cannot cause a large jump}. \rev{If \Healthscore{} exceeds the threshold, the batch processing memory increases to accelerate the learning process; otherwise it decreases.} \rev{For example, if $\alpha = 2$, $\Healthscore{}^{B}_t = 0.6$, and $\Threshold{}_{t} = 0.5$, then the batch processing memory increases by a factor of $1 + 2 \cdot (0.6 - 0.5) = 1.2$. The batch size is $B_t=\lfloor MB_t / M_{batch} \rfloor$, where $M_{batch}$ is the memory consumed per batch sample, estimated by the footprint model described below.}

The replay buffer memory is managed in the same manner:
\begin{equation}
MR_t = \begin{cases}
MR_0, & \text{if } t = 0 \\
MR_{t-1} \cdot \rev{\mathrm{clip}_{[1/s,\, s]}}\!\left(1 + \beta \cdot (\rev{\Healthscore{}^{R}_t} - \Threshold{}_{t})\right), & \text{if } t > 0
\end{cases}
\label{eq:mr}
\end{equation}
\rev{where $\beta$ controls the sensitivity of the replay buffer memory}. \rev{A larger buffer allows more diverse samples during training and improves stability.} \rev{The replay buffer capacity is $R_t=\lfloor MR_t / M_{df} \rfloor$, where $M_{df}$ is the memory of one stored exemplar.} \rev{In our prototype exemplars are materialized in memory (rather than referenced by index into a resident copy of the whole dataset, which an on-device learner does not have), so $M_{df}$ is known exactly from the input shape and shrinking the buffer frees memory immediately. Growing the buffer takes effect for the current and future experiences only, since data of past experiences is no longer available.}

Memory allocation for optimization strategies is managed as follows:
\begin{equation}
MO_t = \begin{cases}
MO_{advanced}, & \text{if } \rev{\Healthscore{}^{O}_t} \geq \Threshold{}_{t} \rev{\text{ and the plugins have not been found harmful}} \\
MO_{default}, & \text{otherwise}
\end{cases}
\label{eq:mo}
\end{equation}
where $MO_{advanced}$ is the memory of the advanced optimization plugins \rev{(GEM and EWC in our prototype)}, which improve the model's ability to learn new knowledge, and $MO_{default}$ that of the default configuration without them. \rev{Instead of a fixed factor relating $MO_{advanced}$ and $MO_{default}$, the footprint model measures the increment of enabling the plugins at runtime. Because the benefit of a plugin depends on the workload (on class-incremental CIFAR, GEM and EWC reduce the accuracy on new classes; on Endless-Sim they slightly improve it, Table~\ref{tab:optimized_ocl}), \Approach{} also checks it online: it keeps the learning accuracy on the new experience plus the change in stability for experiences trained with and without the plugins, and keeps the plugins disabled while the recent average with them is lower, re-trying every five experiences.}

\begin{revblock}
\noindent\textbf{Hard memory-budget constraint.} \Healthscore{} only expresses a preference: its memory term is a soft penalty, and with a low threshold the multiplicative rules above could still request more memory when usage is already at the limit. \Approach{} therefore projects every candidate allocation onto the budget before applying it. It maintains a linear footprint model
\begin{equation}
F(B, R, o) = M_{fix} + a\, B_{eff} + b\, R + o\, M_{opt},
\label{eq:footprint}
\end{equation}
where $B_{eff} = 2B$ once replay is active (each stream mini-batch is paired with a replay mini-batch of the same size), $b = M_{df}$ is the exact size of an exemplar (zero on discrete-GPU servers, where the buffer resides in host memory outside the GPU budget), $a$ is the activation and gradient memory per training sample, calibrated by one forward/backward probe at two batch sizes before the stream starts and refitted from the measured per-experience peaks, $M_{opt}$ is the measured increment of the plugins, and the intercept $M_{fix}$ absorbs any residual so that the model is never optimistic for a configuration that has already been observed. If $F(B_t, R_t, o_t) > (1 - h_{safe}) M_{max}$, \Approach{} first disables the optimization plugins and then shrinks the larger of the two remaining consumers in steps of 20\% until the allocation is feasible. In addition, if the measured peak of the last experience already exceeded $(1 - h_{safe}) M_{max}$, the controller caps both allocations at 70\% of their last value regardless of \Healthscore{}. The scenario in which \Healthscore{} is small but above a decayed threshold while memory is exhausted therefore cannot cause growth: the projection rejects it, and the back-off shrinks the allocation.

\noindent\textbf{Abrupt distribution shifts.} Before training an experience, \Approach{} computes the loss of the current model on a probe of 256 incoming samples. If it exceeds three times the running average observed at previous boundaries, \Approach{} treats the experience as a regime change: it restarts the threshold schedule ($t_0 \leftarrow t$), re-anchors the set-points on the next measurements so that accuracy drops caused by the new domain are not mistaken for a poor configuration, and allows one larger step ($s=4$) so that allocations can be re-sized aggressively instead of incrementally. \Approach{} does not flush the replay buffer upon a shift, because the stored exemplars carry the stability of the previous domain; experience-balanced replacement gives the new domain its share of the buffer from the first experience on.
\end{revblock}

\begin{algorithm}[!tbp]
\footnotesize
\caption{\Healthscore{}-based Self-adaptive Memory Management for OCL \rev{(revised)}}
\label{alg:healthscore_memory}
\begin{revblock}
\begin{algorithmic}[1]
\State \textbf{Input:} OCL stream $e_0, \ldots, e_{T-1}$; memory budget $M_{max}$; initial allocations $MB_0$, $MR_0$, $MO_0$; importance order (weights $w$)
\State Calibrate the footprint model $F$ (Eq.~\ref{eq:footprint}); $\ell^{ref} \gets$ none
\For{$t = 0, \ldots, T-1$}
    \If{$t > 0$} \Comment{measurements of $e_{t-1}$ are available}
        \State Probe $e_t$; if a distribution shift is detected, $t_0 \gets t$ and reset set-points
        \State Compute $\Healthscore{}^{B}_t, \Healthscore{}^{R}_t, \Healthscore{}^{O}_t$ (Eq.~\ref{eq:knob_urge}) and $\Threshold{}_t$ (Eq.~\ref{eq:threshold})
        \State $MB_t \gets MB_{t-1} \cdot \mathrm{clip}(1 + \alpha (\Healthscore{}^{B}_t - \Threshold{}_t))$; \quad $MR_t \gets MR_{t-1} \cdot \mathrm{clip}(1 + \beta (\Healthscore{}^{R}_t - \Threshold{}_t))$
        \State $MO_t \gets MO_{advanced}$ if $\Healthscore{}^{O}_t \geq \Threshold{}_t$ and the plugin check allows it, else $MO_{default}$
        \State If the last peak exceeded $(1-h_{safe})M_{max}$: back off $MB_t, MR_t$ to $70\%$, $MO_t \gets MO_{default}$
        \State Project $(MB_t, MR_t, MO_t)$ onto $F \le (1-h_{safe})M_{max}$
    \EndIf
    \State Reconfigure: $B_t \gets \lfloor MB_t / M_{batch} \rfloor$, $R_t \gets \lfloor MR_t / M_{df} \rfloor$, enable/disable plugins
    \State Train on $e_t$ (one pass); the prefetcher stages stream and replay mini-batches ahead of the GPU
    \State Evaluate; measure $P_t$, $S_t$, $\ell_t$, $M_t$; update the footprint model
\EndFor
\end{algorithmic}
\end{revblock}
\end{algorithm}

\subsection{Automatic Hyperparameter Configuration}
\label{sec:auto_hyper}
\rev{The importance order is turned into the exponents $w_x$ of Eq.~\ref{eq:health_score} by a rule-based method.} Users provide the importance order as a list of metrics, e.g., [memory, plasticity, stability, training latency]. \rev{\Approach{} assigns weights according to the position in the list, with the last factor receiving a weight of 1, the second-to-last a weight of 2, and so forth, }\rev{and normalizes them to sum to one. For the order [memory, plasticity, stability, training latency] this gives $(w_M, w_P, w_S, w_L) = (0.4, 0.3, 0.2, 0.1)$.} \rev{Because the terms are dimensionless, these weights act as intended: a term with twice the weight has twice the influence on $\log \Healthscore{}$ for the same relative deviation. A metric can be excluded by setting its weight to zero, which we use for the per-term ablation in Sec.~\ref{sec:ablation}.} \rev{Computing \Healthscore{} costs $O(1)$ per experience boundary, that is, $O(T)$ for a stream of $T$ experiences.}

\begin{revblock}
\noindent\textbf{Remaining hyperparameters.} Table~\ref{tab:hyperparams} lists every remaining constant. They were fixed once, on a pilot run (SplitCIFAR10, ER, one seed, on a development GPU server that is not one of the evaluation platforms), and used unchanged for every benchmark, algorithm, platform, and memory budget in Sec.~\ref{sec:evaluation}; Sec.~\ref{sec:sensitivity} reports the sensitivity of the results to each of them. \Approach{} therefore removes \emph{per-deployment} tuning: moving to a new device, benchmark, or algorithm requires only the memory budget and, optionally, an importance order. It does not remove the need for sensible defaults, and we do not claim that the defaults are optimal for every stream.
\end{revblock}

\begin{table}[!tbp]
\centering
\caption{\rev{Hyperparameters of \Approach{}. All values are fixed across benchmarks, algorithms, platforms, and budgets.}}
\label{tab:hyperparams}
\begin{revblock}
\resizebox{0.8\textwidth}{!}{
\begin{tabular}{lll}
\toprule
Symbol & Meaning & Value \\
\midrule
$\kappa$ & sigmoid gain on relative deviations (Eq.~\ref{eq:health_score}) & 4 \\
$w$ & importance weights (rank rule) & from the user order; default balanced \\
$P^{ref}, S^{ref}, \ell^{ref}$ & set-points & automatic (best-so-far; first experience) \\
$h_{safe}$ & memory safety margin & 0.15 \\
$\Threshold{}_0$, $\Threshold{}_{min}$, $\delta$ & threshold schedule (Eq.~\ref{eq:threshold}) & 0.6, 0.4, 0.3 \\
$\alpha$, $\beta$ & batch / replay sensitivity (Eqs.~\ref{eq:mb}, \ref{eq:mr}) & 2, 1 \\
$s$ & maximum step per experience (after a shift) & 2 (4) \\
$[B_{min}, B_{max}]$, $[R_{min}, R_{max}]$ & knob ranges & [16, 256], [$10^3$, $10^5$] \\
plugin check & window, retry period & last 3 experiences, 5 experiences \\
shift detector & probe size, loss ratio & 256 samples, 3 \\
$MB_0, MR_0, MO_0$ & initial allocation & $B_0 = 64$, $R_0 = 10^4$, plugins off \\
\bottomrule
\end{tabular}}
\end{revblock}
\end{table}

\subsection{System Prototype Implementation}
\label{sec:prototype_implementation}

Our OCL toolchain is developed as an extension of the Avalanche~\citep{avalanche1} continual learning library, with targeted adaptations for ARM64-based embedded platforms. All system-level optimizations are designed as transparent modules, ensuring full compatibility with the standard Avalanche pipeline and imposing minimal engineering overhead for end users. \rev{These enhancements are evaluated in Sec.~\ref{sec:evaluation}.}
\rev{\noindent\textbf{Data prefetcher.} Standard PyTorch \texttt{DataLoader} workers parallelize the loading of \emph{samples} inside one loader. In replay-based OCL, the training loader of an experience is a composite: Avalanche rebuilds it at every experience from the stream data and one loader per replay group, iterates these sub-loaders in the main thread, concatenates their mini-batches, and only then copies the result to the GPU. This merge and the host-to-device copy remain on the critical path even with many workers. Our prefetcher wraps the composite loader of the current experience (stream-only for the first experience, stream plus replay afterwards), runs the merge and a pinned, non-blocking host-to-device copy on a side CUDA stream in a background thread, and keeps two mini-batches ahead of the training loop. It is not asynchronous loading per se, which DataLoader workers already provide, but a single prefetching path that covers the raw stream and the replay buffer together; it is installed as an Avalanche plugin and is agnostic to the OCL algorithm. Sec.~\ref{sec:ablation} separates its effect from that of DataLoader workers: on the Jetson AGX Orin the latency benefit comes from parallel loading, and the prefetcher adds no measurable speedup on top of the workers, so we regard it as an engineering component of the prototype rather than as a contribution.}
\rev{\noindent\textbf{Memory accounting.} On unified-memory SoCs the budgeted footprint is the process resident set plus the CUDA allocator reserve, since CPU and GPU share DRAM; on discrete-GPU servers it is the CUDA reserve. A watchdog thread samples the footprint every 50\,ms; a run whose footprint exceeds the budget is terminated and reported as OOM. This is how we emulate memory budgets smaller than the physical memory of a board, and it makes the outcome independent of swap configuration.}
\section{Evaluation}
\label{sec:evaluation}

\subsection{Experimental Setup}

\noindent \textbf{Testbeds.} \rev{Our testbeds cover three memory regimes: an edge server with an NVIDIA RTX A4500 GPU (20\,GB), an NVIDIA Jetson AGX Orin (32\,GB of unified memory), and the same Orin board under an emulated 8\,GB budget, which represents the memory class of Jetson Orin NX / Xavier NX modules. Jetson platforms are widely used in autonomous driving~\citep{kato2018autoware,kisavcanin2017deep} and robotics~\citep{popov2022nvradarnet,Duckiebot(DB-J),SparkFun_JetBot,Waveshare_JetBot}. The Jetson AGX Xavier used in the submitted version is no longer available to us; we therefore re-ran every experiment of this revision on the three platforms above, and the Xavier results have been removed. On the Jetson, the budgeted footprint is the process resident set plus the CUDA allocator reserve (unified memory); on the server it is the GPU allocator reserve. A run whose footprint exceeds the budget is terminated and counted as out-of-memory (OOM) (Sec.~\ref{sec:prototype_implementation}); this applies to every method, including \Approach{}. For fixed-configuration baselines, the 8\,GB outcome is obtained from the full-budget Orin run, since the configuration, and hence the memory trajectory, is the same: such a run is OOM under 8\,GB if and only if its measured peak exceeds 8\,GB.}

\noindent \textbf{OCL algorithms and benchmarking dataset.} To ensure a comprehensive evaluation of \Approach{}, we assess the performance metrics on four prominent OCL algorithms, including ER~\citep{chaudhry2019continual}, GSS~\citep{aljundi2019gradient}, GEM~\citep{lopez2017gradient}, and AGEM~\citep{chaudhry2018efficient}, based on ResNet-20~\citep{he2016deep} DNN architecture. These algorithms represent various OCL approaches, ensuring a comprehensive assessment of \Approach{} across various settings.  We evaluate \Approach{} on a set of widely studied OCL benchmarks involving different sizes and different OCL tasks as detailed in Tab.~\ref{tab:models}.

\begin{table}[!tbp]
\centering
\caption{Statistics of benchmark datasets used in this work. IC: Incremental Class, IL: Incremental Illumination, WC: Weather Change, NI: New Instances, NC: New Classes, NIC: New Instances and Classes.}
\resizebox{0.7\textwidth}{!}{
\begin{tabular}{lccccc}
\toprule
Dataset & Size & Task & \# Experience & \# Images & Image Size \\
\midrule
SplitCIFAR10~\citep{krizhevsky2009learning} & Small & IC & 10 & 60,000 & 32$\times$32 \\
SplitCIFAR100~\citep{krizhevsky2009learning} & Small & IC & 10 & 60,000 & 32$\times$32 \\
\midrule
\multirow{3}{*}{CORe50~\citep{lomonaco2019fine,lomanco2017core50}} & Medium & NI & 8 & 164,866  & 32$\times$32 \\
& Medium & NC & 9 & 164,866  & 32$\times$32 \\
& Large & NIC & 79 & 164,866  & 32$\times$32 \\
\midrule
\midrule
\multirow{3}{*}{\texttt{Endless-Sim}~\citep{hess2021procedural}} & N/A & IC & 4 & 49,134 & 32$\times$32 \\
& N/A & IL & 5 & 181,899 & 32$\times$32 \\
& N/A & WC & 5 & 188,536 & 32$\times$32 \\
\bottomrule
\end{tabular}
}
\label{tab:models}

\end{table}
.

\noindent \textbf{Robotic case study.} To further demonstrate the practicality of \Approach{} in realistic scenarios, we conduct a robotic case study in the context of autonomous driving using \texttt{Endless-Sim}~\citep{hess2021procedural}, under incremental class learning (IC), incremental illumination conditions (IL), and weather changes (WC).

\noindent\textbf{Baselines.}
We compare \Approach{} to the following approaches:
\begin{itemize}[leftmargin=10pt]
\item \textbf{LR~\citep{pellegrini2020latent}:} A state-of-the-art baseline leveraging latent replay for real-time continual learning in robotics. \rev{After the first experience, the layers below the latent layer (stem and the first two ResNet stages) are frozen, so replayed samples only update the upper part of the network, which reduces the backward computation.} It uses the default \texttt{Avalanche-lib} parameters.
\item \textbf{MAX-A~\citep{avalanche1}}: MAX-A uses preset values for training parameters in \texttt{Avalanche-lib} for \rev{empirically} good plasticity and stability in general OCL scenarios \rev{(batch size 16, replay capacity $10^6$, GEM and EWC plugins enabled)}.
\item \textbf{MAX-P~\citep{avalanche1}}: \rev{a high-efficiency configuration of \texttt{Avalanche-lib}, targeting training latency and throughput with a large batch size} \rev{(batch size 1024, replay capacity 2048, no plugins)}.
\item \rev{\textbf{Miro~\citep{ma2023cost}}: the closest system-level competitor (Sec.~\ref{sec:related}). We re-implemented Miro's controller in our pipeline: at every experience from the second one on, it samples ten feasible candidates from a grid of batch sizes $\{16,\ldots,256\}$ and replay capacities $\{10^3,\ldots,10^5\}$, trains a fork of the current model on a 5\% slice of the incoming data for each candidate, keeps the top 50\% by accuracy on a held-out slice of buffered and incoming data, and selects the candidate with the highest accuracy per unit of predicted energy. Profiling time and energy are charged to Miro. Deviations from the original, required by single-pass OCL and our devices, are listed in Appendix~\ref{app:miro}.}
\end{itemize}
\rev{For the ablation of the adaptive mechanism (Sec.~\ref{sec:ablation}) we further use \textbf{Static-Init} (\Approach{}'s initial configuration kept fixed), \textbf{Static-Tuned} (the most accurate configuration of a $3\times3$ grid of batch sizes $\{16, 64, 256\}$ and replay capacities $\{10^3, 10^4, 10^5\}$, selected offline on seed 0 and then run as a fixed configuration; nine offline runs per benchmark), and \textbf{Heuristic}, a memory-only controller that grows batch size and replay capacity by 25\% while the measured peak is below 60\% of the budget and halves them above 85\%. All three use exactly the same data-loading path (DataLoader workers and the prefetcher) as \Approach{}.}

\noindent\textbf{Implementation details.} \rev{All methods share the same pipeline: ResNet-20, Adam with learning rate $10^{-3}$, one pass per experience, and a replay buffer whose exemplars are stored in memory (Sec.~\ref{sec:fine_grained_control}); stored exemplars are re-augmented with the benchmark's own training augmentation when replayed.} \rev{Every result is the mean $\pm$ standard deviation over three seeds (0, 1, 2), which change the class order, the initialization, and the data order. Plasticity is the mean accuracy on each experience right after learning it; stability is one minus the mean forgetting of previous experiences at the end of the stream (Sec.~\ref{sec:background_OCL}); training latency is the wall-clock time spent in training, excluding evaluation, and including Miro's profiling and \Approach{}'s controller.}

\begin{figure}[!tbp]
    \centering
    \includegraphics[width=\textwidth]{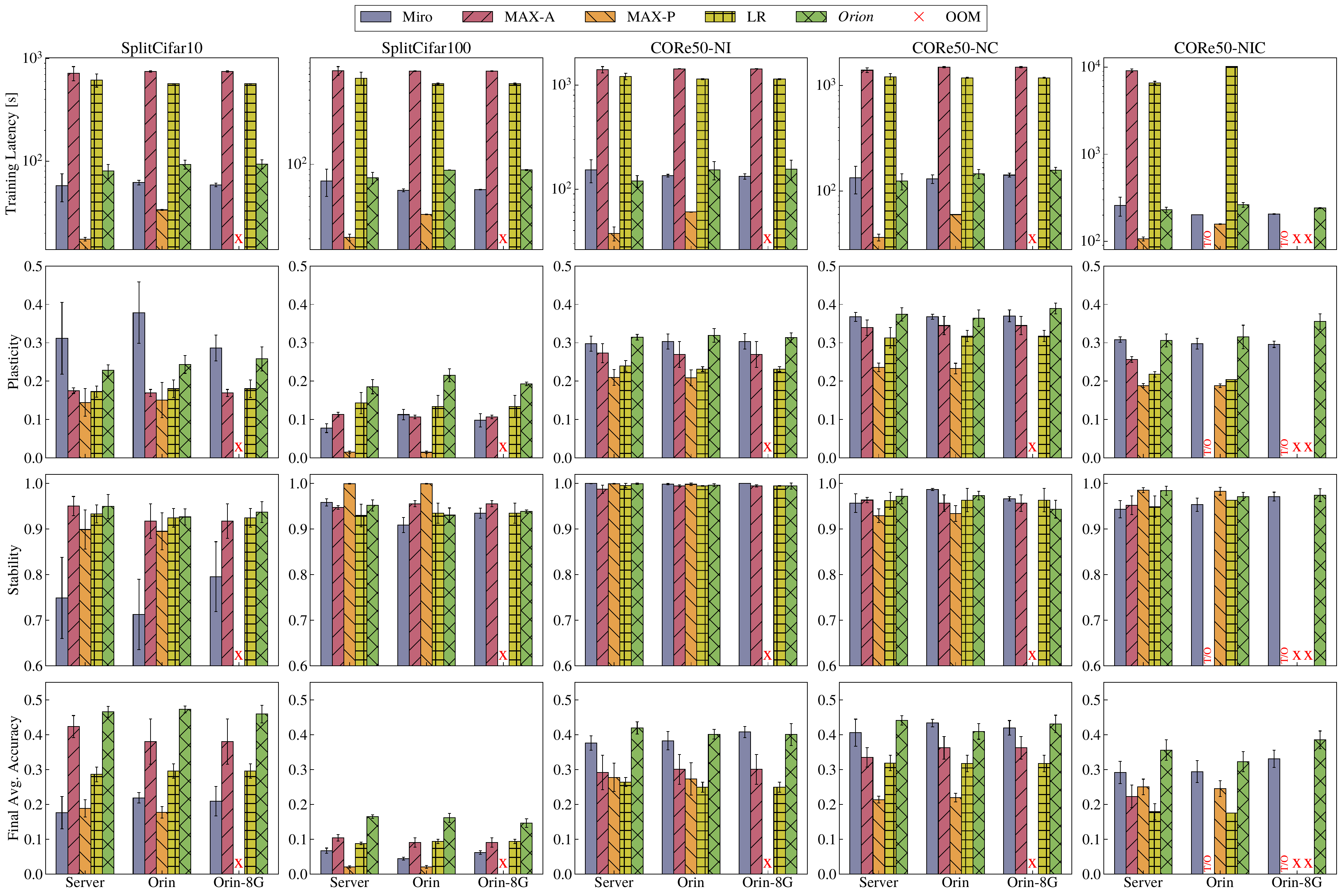}
    \caption{\rev{Overall effectiveness with ER on five benchmarks on the edge server, the Jetson AGX Orin, and the Orin under an 8\,GB budget (mean and standard deviation over three seeds; latency on a log scale). Orin-8G: the Orin under an 8\,GB budget. Red $\times$: out of memory in all seeds; T/O: did not finish within 4\,h. Full numbers in Tables~\ref{tab:main_server_bench}--\ref{tab:main_orin8_bench} (Appendix~\ref{app:tables}).}}
    \label{fig:overall_effectiveness_benchmark}
    
\end{figure}

\begin{figure}[!tbp]
    \centering
    \includegraphics[width=\textwidth]{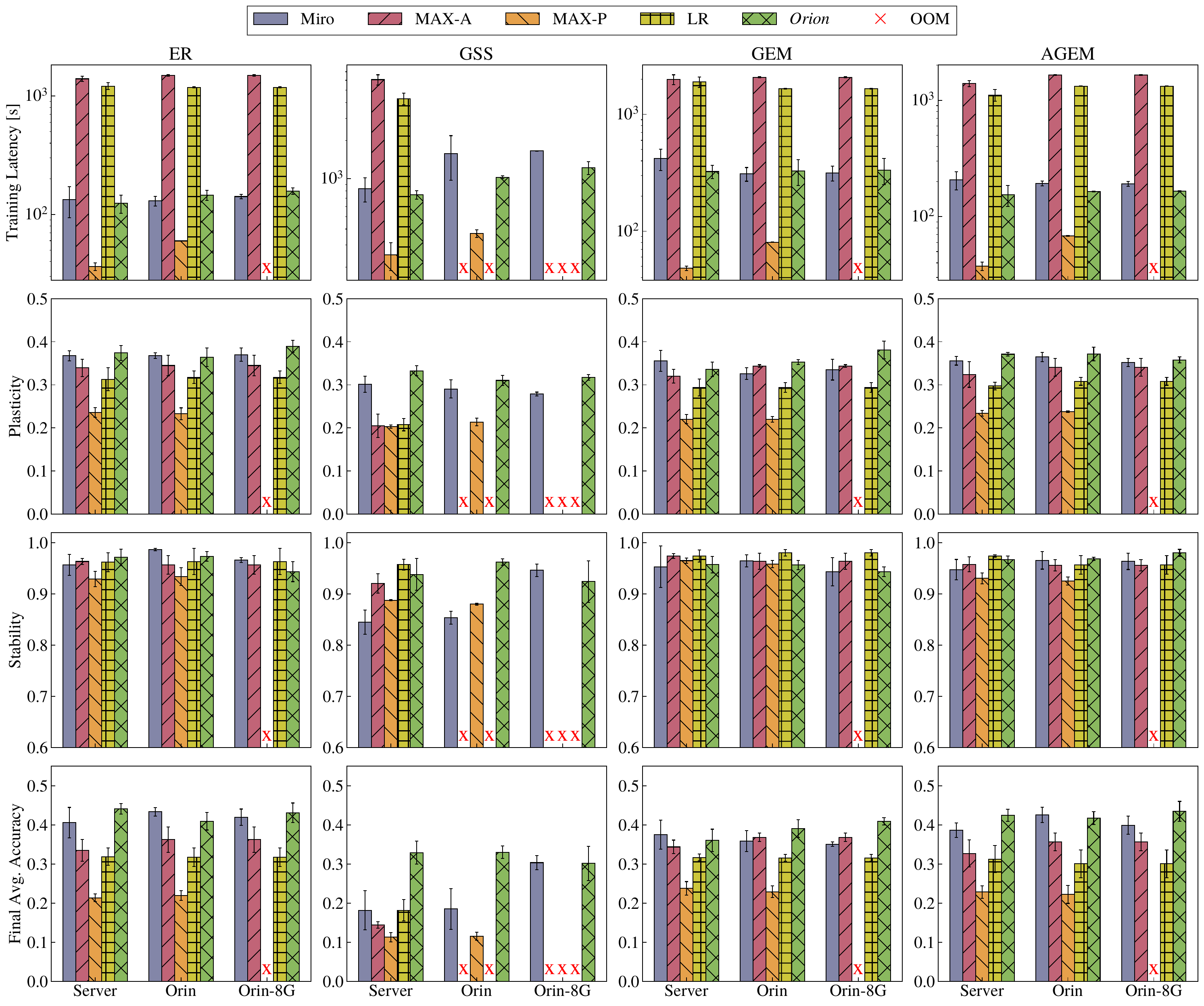}
    \caption{\rev{Overall effectiveness of four OCL algorithms on CORe50-NC (mean and standard deviation over three seeds). Notation as in Fig.~\ref{fig:overall_effectiveness_benchmark}; full numbers in Tables~\ref{tab:main_server_algo}--\ref{tab:main_orin8_algo}.}}
    
    \label{fig:overall_effectiveness_algorithm}
\end{figure}

\subsection{Overall Effectiveness}
\label{sec:overall_effectiveness}

\begin{revblock}
This section compares \Approach{} with the baselines on the three platforms (Figs.~\ref{fig:overall_effectiveness_benchmark} and~\ref{fig:overall_effectiveness_algorithm}). Following the reviewers' advice, we separate two questions: whether a method can run at all under a device's memory budget (\emph{feasibility}), and how methods compare when they all run (\emph{like-for-like quality and latency}). Every number is the mean over three seeds; the standard deviations are given in the figures and in Appendix~\ref{app:tables}. Besides plasticity and stability, we report the final average accuracy over all experiences (ACC), the most common summary metric in continual learning, because plasticity and stability alone can be misleading when a method learns little (a model that learns nothing forgets nothing).

\noindent\textbf{Feasibility under memory budgets.} The fixed-configuration baselines fail as soon as their configuration does not fit. On the 32\,GB Orin, MAX-A and LR run out of memory with GSS in all seeds, because GSS keeps its $10^6$-sample memory on the GPU, and MAX-A does not finish CORe50-NIC within 4\,h. Under the 8\,GB budget, MAX-P runs out of memory on every benchmark and with every algorithm (its footprint reaches 9.6--11.2\,GB with a batch of 1024 and replay), and LR on CORe50-NIC and with GSS. Miro runs out of memory with GSS in one of three seeds under 8\,GB, because its trial forks add to GSS's device-resident memory. \Approach{} completes every run on every platform; on CORe50-NIC, where its footprint reaches 8.9\,GB on the unconstrained Orin, the hard constraint keeps it at 7.1\,GB under the 8\,GB budget, and its accuracy is even higher (0.385 vs.\ 0.323) because the projection also caps the batch growth discussed in Sec.~\ref{sec:ablation}. On the edge server (20\,GB of GPU memory) all methods complete.

\noindent\textbf{Like-for-like comparison with ER.} On the edge server, where all methods complete, \Approach{} reaches the highest final accuracy on all five benchmarks: on average 9.4 points above MAX-A (range $+4.3$ to $+13.3$), 14.2 points above LR, 17.9 points above MAX-P, and 10.6 points above Miro. Its plasticity is higher than that of MAX-A, LR, and MAX-P on every benchmark (by 5.0 and 6.5 points over MAX-A and LR on average), and its stability is at least as high as theirs except against MAX-P on SplitCIFAR100 ($-4.7$ points), where MAX-P learns almost nothing (plasticity 0.014) and therefore forgets almost nothing. It trains 16.3$\times$ faster than MAX-A (range 9.0--39.6$\times$) and 13.0$\times$ faster than LR, whereas MAX-P, which only optimizes throughput, is about $3\times$ faster than \Approach{} but reaches by far the lowest accuracy. On the Jetson AGX Orin the picture is the same against the fixed baselines: \Approach{} is 9.0$\times$ faster than MAX-A and 13.4$\times$ faster than LR and more accurate than both on every benchmark (by 7.8 and 12.7 points on average). Against Miro, \Approach{} is more accurate on four of the five benchmarks on the Orin (by up to 25.4 points on SplitCIFAR10, where Miro's higher plasticity comes with a 21-point lower stability), while Miro is 2.4 points more accurate on CORe50-NC; on this board Miro's total latency, including profiling, is 23\% lower than \Approach{}'s on average. Sec.~\ref{sec:miro_comparison} discusses the comparison with Miro in detail.

\noindent\textbf{Across OCL algorithms.} With GEM, AGEM, and GSS on CORe50-NC (Fig.~\ref{fig:overall_effectiveness_algorithm}), \Approach{} is the most accurate method with GSS on both platforms (0.329 on the server and 0.330 on the Orin, versus at most 0.185 for any baseline) and with GEM on the Orin, the most accurate with AGEM on the server, and within 1.5 points of the best with GEM on the server and AGEM on the Orin, where Miro is ahead. In all cases it is 5--11$\times$ faster than MAX-A and LR. GSS shows most clearly why a fixed configuration is brittle: MAX-A's configuration is the most accurate fixed choice with ER, yet it does not fit the Orin with GSS at all.

\noindent\textbf{Comparison with the submitted results.} The revised experiments replace the submitted Figures~4--5 and Tables~3--4 (Sec.~\ref{sec:evaluation}). The qualitative conclusions of the submitted version hold: \Approach{} avoids the OOM failures of the fixed baselines and is an order of magnitude faster than MAX-A and LR. The quantitative ones change. The speedup over MAX-A is 9--16$\times$ depending on the platform (the submitted version reported 12.13$\times$), and \Approach{} now also \emph{improves} plasticity and final accuracy over the baselines instead of trading them off. The comparison with the offline Oracle has been removed (see Sec.~\ref{sec:ablation} for an offline-tuned static baseline): its 392$\times$ and 267$\times$ ``speedups'' compared a single online run with 42 offline runs and were not per-method latency comparisons.
\end{revblock}

\subsubsection{Performance across user-specific preferences.}
\begin{table}[!tbp]
\centering\ifrevhighlight\color{blue}\fi
\caption{\Approach{} under different importance orders (ER, Jetson AGX Orin, seed 0). Prefer Latency: [latency, memory, plasticity, stability]; Balanced: [memory, plasticity, stability, latency]; Prefer P/S: [plasticity, stability, memory, latency].}
\label{tab:user_preferences}
\resizebox{0.95\textwidth}{!}{
\begin{tabular}{l|ccc|ccc|ccc|ccc}
\toprule
 & \multicolumn{3}{c|}{Latency [s] $\downarrow$} & \multicolumn{3}{c|}{Plasticity $\uparrow$} & \multicolumn{3}{c|}{Stability $\uparrow$} & \multicolumn{3}{c}{Final avg.\ acc.\ $\uparrow$} \\
Benchmark & Lat. & Bal. & P/S & Lat. & Bal. & P/S & Lat. & Bal. & P/S & Lat. & Bal. & P/S \\
\midrule
SplitCIFAR10 & 88 & 87 & 89 & 0.296 & 0.235 & 0.287 & 0.903 & 0.922 & 0.921 & 0.440 & 0.459 & 0.470 \\
SplitCIFAR100 & 88 & 88 & 91 & 0.199 & 0.238 & 0.204 & 0.926 & 0.908 & 0.933 & 0.133 & 0.160 & 0.157 \\
CORe50-NI & 172 & 133 & 138 & 0.328 & 0.331 & 0.326 & 0.998 & 0.998 & 0.994 & 0.408 & 0.415 & 0.402 \\
CORe50-NC & 157 & 166 & 144 & 0.368 & 0.341 & 0.387 & 0.963 & 0.961 & 0.962 & 0.428 & 0.381 & 0.431 \\
CORe50-NIC & 249 & 257 & 250 & 0.369 & 0.286 & 0.339 & 0.938 & 0.980 & 0.925 & 0.331 & 0.313 & 0.292 \\
\bottomrule
\end{tabular}}
\end{table}

\begin{revblock}
Table~\ref{tab:user_preferences} runs \Approach{} with three importance orders on the Jetson AGX Orin. In the revised controller, the order has only a small and inconsistent effect. Preferring latency reduces the training latency on none of the five benchmarks by more than 6\% relative to the balanced order, and it is slower on CORe50-NI (172\,s vs.\ 133\,s); preferring plasticity and stability gives the highest final accuracy on SplitCIFAR10 and CORe50-NC but not on the other three benchmarks, and the differences are mostly within the seed-to-seed variation of Table~\ref{tab:main_orin_bench}. The reason is that, with dimensionless terms and per-knob indicators, the weights only rescale terms that already point in the same direction for a given knob: the batch indicator is dominated by the plasticity deficit under every order, and the replay indicator does not contain the latency term at all. Unlike the submitted version, we therefore do not claim that the importance order yields a controllable trade-off; making it effective, for instance by mapping a latency preference to a latency target $\ell^{ref}$, is future work (Sec.~\ref{sec:discussion}).
\end{revblock}

\noindent\begin{minipage}{\linewidth}
\begin{shaded}
\noindent\textbf{Overall effectiveness and versatility}: \rev{\Approach{} completes every run under every memory budget, is an order of magnitude faster than MAX-A and LR, and reaches the highest final accuracy in most settings; where all methods fit, the gains come from better configurations rather than from the baselines' OOM failures.}
\end{shaded}
\end{minipage}

\begin{revblock}
\subsection{Comparison with Miro}
\label{sec:miro_comparison}

Miro~\citep{ma2023cost} is the closest system to \Approach{} (Sec.~\ref{sec:related}), and we compare the two directly with the same pipeline, data loading, and memory budgets (Tables~\ref{tab:main_server_bench} and~\ref{tab:main_server_algo}; Miro's profiling time is included in its training latency). \Approach{} reaches a higher final average accuracy on all five benchmarks with ER: 0.466 vs.\ 0.176 on SplitCIFAR10, 0.165 vs.\ 0.067 on SplitCIFAR100, 0.420 vs.\ 0.376 on CORe50-NI, 0.441 vs.\ 0.406 on CORe50-NC, and 0.356 vs.\ 0.291 on CORe50-NIC (edge server, three seeds), and also with AGEM (0.425 vs.\ 0.387); with GEM, Miro is slightly better (0.375 vs.\ 0.360). The total training latency of the two methods is of the same order: Miro trains faster, but its per-experience profiling (ten trial forks on a 5\% slice) costs between 35\,s (SplitCIFAR10) and 152\,s (CORe50-NIC), so that its total latency is lower than \Approach{}'s on the two CIFAR benchmarks (58\,s vs.\ 80\,s and 70\,s vs.\ 75\,s) and higher on the CORe50 benchmarks (e.g., 258\,s vs.\ 231\,s on CORe50-NIC). The accuracy gap has a simple cause. Miro's single-pass trials rank candidates by their accuracy after training on 5\% of an experience; in class-incremental streams this short-horizon signal favours configurations that fit the new classes quickly, such as large batches (Miro selects a batch size of at least 128 in 83\% and 92\% of its decisions on SplitCIFAR10 and SplitCIFAR100), at the cost of forgetting and of fewer updates per pass, which the short trials cannot observe. \Approach{} instead reacts to the stability and plasticity measured over the whole stream. Miro's design was developed for multi-epoch tasks, where a trial of a few epochs is a much better predictor of the outcome; our results do not contradict its findings in that setting, but show that per-experience profiling transfers poorly to single-pass online continual learning.
\end{revblock}

\begin{table*}[!tbp]
\centering\ifrevhighlight\color{blue}\fi
\caption{Robotic case study on Jetson AGX Orin (Endless-Sim; IC: incremental classes, IL: incremental illumination, WC: weather changes; one seed). \ding{55}: out of memory on the 32\,GB board. Best value per algorithm and row in bold.}
\label{tab:robotic_case_results}
\resizebox{\textwidth}{!}{
\begin{tabular}{ll|ccccc|ccccc|ccccc|ccccc}
\toprule
 & Algorithm & \multicolumn{5}{c|}{ER} & \multicolumn{5}{c|}{GSS} & \multicolumn{5}{c|}{GEM} & \multicolumn{5}{c}{AGEM} \\
Metric & Scenario & MAX-A & MAX-P & LR & Miro & \Approach{} & MAX-A & MAX-P & LR & Miro & \Approach{} & MAX-A & MAX-P & LR & Miro & \Approach{} & MAX-A & MAX-P & LR & Miro & \Approach{} \\
\midrule
\multirow{3}{*}{Latency [s] $\downarrow$} & IC & 244 & \ding{55} & 165 & 70 & \textbf{61} & \ding{55} & \ding{55} & \ding{55} & 862 & \textbf{567} & 281 & \ding{55} & 227 & 160 & \textbf{112} & 247 & \ding{55} & 194 & 115 & \textbf{75} \\
 & IL & 772 & \ding{55} & 597 & 215 & \textbf{171} & \ding{55} & \ding{55} & \ding{55} & 1702 & \textbf{1468} & 1017 & \ding{55} & 822 & \textbf{273} & 289 & 874 & \ding{55} & 698 & 253 & \textbf{206} \\
 & WC & 798 & \ding{55} & 624 & 239 & \textbf{198} & \ding{55} & \ding{55} & \ding{55} & 2241 & \textbf{1304} & 1063 & \ding{55} & 835 & 341 & \textbf{287} & 898 & \ding{55} & 738 & 253 & \textbf{242} \\
\midrule
\multirow{3}{*}{Plasticity $\uparrow$} & IC & 0.99 & \ding{55} & 0.98 & 0.99 & \textbf{1.00} & \ding{55} & \ding{55} & \ding{55} & 0.70 & \textbf{0.82} & 0.99 & \ding{55} & 0.99 & \textbf{1.00} & 0.93 & 0.99 & \ding{55} & 0.97 & \textbf{1.00} & 0.99 \\
 & IL & 0.95 & \ding{55} & 0.94 & 0.98 & \textbf{0.99} & \ding{55} & \ding{55} & \ding{55} & 0.90 & \textbf{0.97} & 0.95 & \ding{55} & 0.93 & \textbf{0.96} & 0.95 & 0.94 & \ding{55} & 0.97 & \textbf{0.98} & 0.98 \\
 & WC & 0.93 & \ding{55} & 0.93 & \textbf{0.96} & 0.93 & \ding{55} & \ding{55} & \ding{55} & 0.67 & \textbf{0.89} & 0.91 & \ding{55} & 0.92 & 0.95 & \textbf{0.96} & 0.92 & \ding{55} & 0.83 & \textbf{0.95} & 0.91 \\
\midrule
\multirow{3}{*}{Stability $\uparrow$} & IC & 0.94 & \ding{55} & 0.83 & 0.93 & \textbf{0.99} & \ding{55} & \ding{55} & \ding{55} & 0.80 & \textbf{0.99} & \textbf{0.97} & \ding{55} & 0.87 & 0.96 & 0.91 & \textbf{0.95} & \ding{55} & 0.83 & 0.91 & 0.91 \\
 & IL & 0.99 & \ding{55} & 0.98 & 0.99 & \textbf{0.99} & \ding{55} & \ding{55} & \ding{55} & 0.57 & \textbf{1.00} & 0.96 & \ding{55} & 0.99 & \textbf{0.99} & 0.96 & 1.00 & \ding{55} & 0.99 & 0.99 & \textbf{1.00} \\
 & WC & 1.00 & \ding{55} & 0.98 & \textbf{1.00} & 1.00 & \ding{55} & \ding{55} & \ding{55} & 0.35 & \textbf{1.00} & \textbf{1.00} & \ding{55} & 0.96 & 0.99 & 0.97 & 1.00 & \ding{55} & 0.99 & 1.00 & \textbf{1.00} \\
\midrule
\multirow{3}{*}{Final avg.\ acc.\ $\uparrow$} & IC & 0.94 & \ding{55} & 0.86 & 0.94 & \textbf{0.99} & \ding{55} & \ding{55} & \ding{55} & 0.56 & \textbf{0.98} & 0.97 & \ding{55} & 0.89 & \textbf{0.97} & 0.91 & \textbf{0.96} & \ding{55} & 0.84 & 0.93 & 0.93 \\
 & IL & 0.94 & \ding{55} & 0.93 & 0.98 & \textbf{0.99} & \ding{55} & \ding{55} & \ding{55} & 0.56 & \textbf{0.98} & 0.92 & \ding{55} & 0.94 & \textbf{0.97} & 0.92 & 0.98 & \ding{55} & 0.97 & 0.98 & \textbf{0.99} \\
 & WC & 0.93 & \ding{55} & 0.93 & \textbf{0.99} & 0.98 & \ding{55} & \ding{55} & \ding{55} & 0.15 & \textbf{0.95} & \textbf{0.96} & \ding{55} & 0.89 & 0.95 & 0.94 & 0.97 & \ding{55} & 0.95 & \textbf{0.98} & 0.98 \\
\bottomrule
\end{tabular}}
\end{table*}

\subsection{Robotic Case Study}
\label{sec:oclrobot_case_study}

\begin{figure}[!tbp]
\centering
    \includegraphics[width=0.9\textwidth]{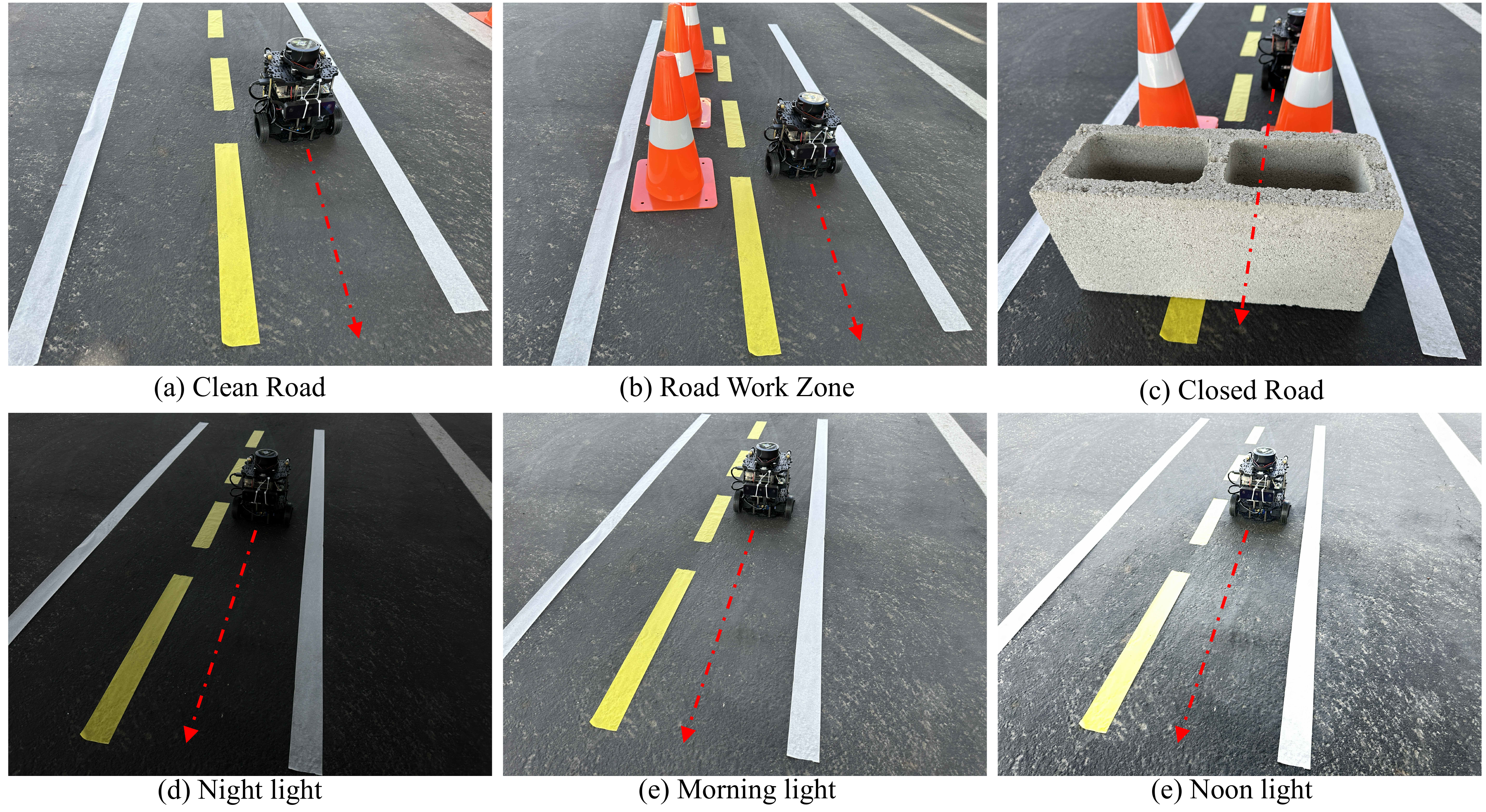}
    \caption{A realistic case study based on an NVIDIA Jetson GPU-enabled autonomous navigation robot built on Turtlebot3. (a) (b) (c) exhibit incremental class scenarios (IC), (d) (e) (f) exhibit incremental light scenarios (IL). Red lines indicate the robot's moving directions.}
    \label{fig:robot-photo}
    
\end{figure}

To evaluate the practicality of \Approach{}, we conducted a case study using a Jetson-enabled autonomous navigation robot, built on the Turtlebot3 Burger~\citep{robotisemanual2024} with an NVIDIA Jetson motherboard and a high-resolution camera for real-time data processing and navigation (Fig.~\ref{fig:robot-photo}). 
\rev{Using \texttt{Endless-Sim}~\citep{hess2021procedural}, we evaluate the methods in the three OCL scenarios of the robot (incremental classes, IC; incremental illumination, IL; weather changes, WC) with all four algorithms on the Jetson AGX Orin (Table~\ref{tab:robotic_case_results}). The images are $64\times64$, so the memory pressure is higher than on the benchmarks of Sec.~\ref{sec:overall_effectiveness}. \Approach{} completes all twelve algorithm--scenario combinations. MAX-P runs out of memory in all twelve on the 32\,GB board (a batch of 1024 images of this size, plus replay, does not fit), and MAX-A and LR run out of memory with GSS in all three scenarios, because GSS allocates its $10^6$-sample memory on the GPU up front. \Approach{} is the fastest method in eleven of the twelve combinations, and its final accuracy is the best in six and within 2 points of the best in three more (ER, GEM, and AGEM in WC); the exceptions are GEM in IC and IL, where it is 6.0 and 5.6 points below Miro, and AGEM in IC, where it is 3.0 points below MAX-A. With GSS, the only method besides \Approach{} that fits is Miro, whose accuracy collapses in IL and WC (0.559 and 0.149 versus 0.981 and 0.948 for \Approach{}).}

\begin{figure}[!t]
\centering
\includegraphics[width=\textwidth]{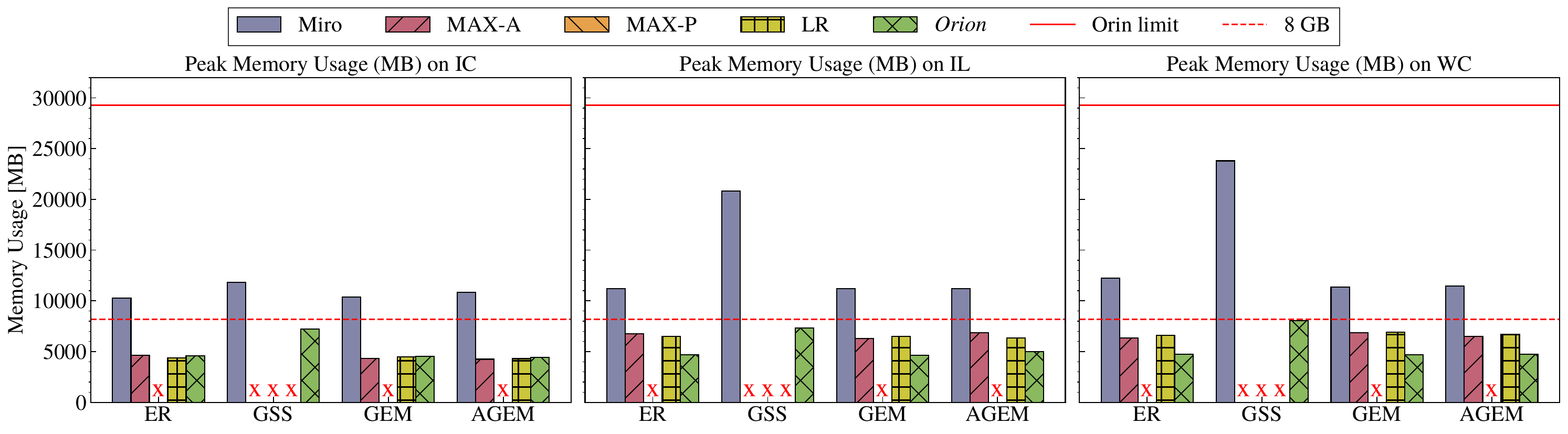}
\caption{\rev{Peak memory usage in the robotic case study on the Jetson AGX Orin. Red $\times$: out of memory. The solid red line marks the usable memory of the 32\,GB board, the dashed red line an 8\,GB device.}}
\label{fig:memory_breakdown_donkeycar}
\end{figure}

\rev{Fig.~\ref{fig:memory_breakdown_donkeycar} explains these outcomes through the peak footprint of each run. MAX-A's and LR's footprints are moderate (4.1--6.8\,GB) as long as the replay memory stays in host DRAM, but GSS moves it to the GPU and the allocation fails. Miro's profiling forks raise its footprint to 10.0--11.9\,GB with ER, GEM, and AGEM and to 11.6--23.2\,GB with GSS, i.e., beyond an 8\,GB device in every case. \Approach{} stays between 4.3 and 7.8\,GB.}

\rev{\noindent\textbf{Stability with ER in IC.} In the submitted version, \Approach{}'s stability with ER in the IC scenario dropped from 0.92 (MAX-A) to 0.36 (R3-W4). With the revised controller this degradation no longer occurs: \Approach{} reaches a stability of 0.994 versus 0.940 for MAX-A, at a quarter of the latency (61\,s vs.\ 244\,s). Two design changes are responsible. The stability term now compares the stability with the best value reached so far on a dimensionless scale, so a drop of a few points already raises the replay indicator, whereas the submitted indicator applied a fixed unit-scale gain to the raw difference; and the replay capacity is driven by its own indicator (Eq.~\ref{eq:knob_urge}), which responds to stability deficits and is not pulled down by the latency term, as the single scalar indicator was.}

\noindent\begin{minipage}{\linewidth}
\begin{shaded}
\noindent\textbf{Practical usability}: \rev{In the robotic scenarios, \Approach{} completes every run within the memory of the board, is the fastest method in eleven of twelve settings, and avoids the stability degradation of the submitted version.}
\end{shaded}
\end{minipage}

\subsection{Ablation Study}
\label{sec:ablation}

\begin{revblock}
\noindent\textbf{Isolating the adaptive mechanism.} Table~\ref{tab:ablation} isolates the contribution of \Healthscore{}-driven adaptation. All variants use exactly the same data-loading path (DataLoader workers and our prefetcher), so differences stem only from the configuration policy.
\begin{itemize}[leftmargin=10pt]
\item \emph{Adaptation versus the same configuration kept fixed.} Static-Init, i.e., \Approach{}'s initial configuration without adaptation, reaches 0.085, 0.378, and 0.358 on SplitCIFAR100, CORe50-NC, and CORe50-NIC, versus 0.165, 0.441, and 0.356 for \Approach{}. Adaptation thus nearly doubles the accuracy on SplitCIFAR100 and adds 6.3 points on CORe50-NC, at a higher training latency (the controller moves towards smaller batches). On CORe50-NIC it brings no gain.
\item \emph{Adaptation versus offline tuning.} Static-Tuned, the best fixed configuration of a $3\times3$ grid selected on seed 0, matches \Approach{} on SplitCIFAR100 (0.163) and is better on CORe50-NC (0.473) and CORe50-NIC (0.446). It requires nine complete runs before deployment, whereas \Approach{} runs once without offline tuning. \Approach{} thus closes the entire gap between Static-Init and Static-Tuned on SplitCIFAR100 and 66\% of it on CORe50-NC, but none of it on CORe50-NIC. Offline tuning remains the higher ceiling when its cost can be paid; \Approach{} is the alternative when it cannot.
\item \emph{A memory-only controller is not sufficient.} The Heuristic baseline grows batch size and buffer while memory is available and shrinks them otherwise. It reaches only 0.019, 0.332, and 0.193, below Static-Init, because it uses memory without regard to what the memory buys (large batches lower plasticity).
\item \emph{Per-knob versus scalar \Healthscore{}.} Driving all knobs with the single scalar \Healthscore{} of the submitted version reaches 0.090, 0.417, and 0.347; the per-knob indicators of Eq.~\ref{eq:knob_urge} improve on it on all three benchmarks.
\item \emph{Individual terms.} Removing the latency term hurts most (0.134, 0.403, 0.267): without it, the batch size is driven down to the minimum and latency rises sharply. Removing the stability term costs 6.7 points on CORe50-NIC, the longest stream, and at most 1.6 points on the shorter ones. Removing the plasticity term costs 2.4 points on SplitCIFAR100 and removing the memory term 1.0 and 2.2 points on SplitCIFAR100 and CORe50-NIC; on the remaining benchmarks these two terms change the result by less than one standard deviation. Every term matters on at least one benchmark.
\item \emph{Threshold decay.} The time-dependent threshold does not help: without it, accuracy is unchanged on SplitCIFAR100 and CORe50-NC (0.170, 0.440) and clearly higher on CORe50-NIC (0.424). On a 79-experience stream, the threshold reaches its floor early, after which the controller keeps growing the allocations (the average batch size is 195 on CORe50-NIC, versus 24 and 35 on SplitCIFAR100 and CORe50-NC), which lowers plasticity; we keep the default of the submitted design for consistency and report this as a limitation (Sec.~\ref{sec:discussion}).
\item \emph{Shift handling.} Disabling the shift detector lowers accuracy on all three benchmarks (0.146, 0.419, 0.295), most on CORe50-NIC, whose experiences introduce new instances and classes in an interleaved order.
\end{itemize}
\end{revblock}

\begin{table*}[!tbp]
\centering\ifrevhighlight\color{blue}\fi
\caption{Ablation of the adaptive mechanism (ER, edge server, mean $\pm$ std over three seeds). All variants use the same data-loading path as \Approach{}. Static-Tuned is selected from a $3\times3$ grid (batch $\{16,64,256\}$ $\times$ replay $\{10^3,10^4,10^5\}$) on seed 0, i.e., it requires nine additional offline runs per benchmark.}
\label{tab:ablation}
\resizebox{\textwidth}{!}{
\begin{tabular}{lcccccccccccc}
\toprule
 & \multicolumn{4}{c}{S-CIFAR100} & \multicolumn{4}{c}{CORe50-NC} & \multicolumn{4}{c}{CORe50-NIC} \\
\cmidrule(lr){2-5} \cmidrule(lr){6-9} \cmidrule(lr){10-13}
Variant & Lat.\ [s] & P & S & ACC & Lat.\ [s] & P & S & ACC & Lat.\ [s] & P & S & ACC \\
\midrule
Static-Init (\Approach{}'s initial configuration) & 22$\pm$1 & 0.21$\pm$0.05 & 0.85$\pm$0.06 & 0.085$\pm$0.007 & 49$\pm$1 & 0.35$\pm$0.00 & 0.97$\pm$0.01 & 0.378$\pm$0.021 & 79$\pm$2 & 0.37$\pm$0.01 & 0.95$\pm$0.01 & 0.358$\pm$0.008 \\
Static-Tuned (best of 9 offline runs) & 62$\pm$3 & 0.20$\pm$0.03 & 0.93$\pm$0.01 & 0.163$\pm$0.007 & 135$\pm$2 & 0.40$\pm$0.01 & 0.98$\pm$0.01 & 0.474$\pm$0.020 & 268$\pm$3 & 0.38$\pm$0.01 & 0.99$\pm$0.00 & 0.446$\pm$0.011 \\
Heuristic (memory-only controller) & 58$\pm$0 & 0.08$\pm$0.01 & 0.92$\pm$0.01 & 0.019$\pm$0.006 & 115$\pm$1 & 0.31$\pm$0.02 & 0.97$\pm$0.00 & 0.333$\pm$0.019 & 655$\pm$4 & 0.22$\pm$0.01 & 0.96$\pm$0.01 & 0.194$\pm$0.014 \\
Scalar \Healthscore{} (submitted version) & 29$\pm$1 & 0.25$\pm$0.02 & 0.82$\pm$0.01 & 0.091$\pm$0.013 & 82$\pm$4 & 0.37$\pm$0.00 & 0.97$\pm$0.01 & 0.417$\pm$0.001 & 174$\pm$66 & 0.31$\pm$0.01 & 0.98$\pm$0.01 & 0.348$\pm$0.023 \\
\Approach{} (full, per-knob \Healthscore{}) & 75$\pm$9 & 0.19$\pm$0.02 & 0.95$\pm$0.01 & 0.165$\pm$0.005 & 124$\pm$22 & 0.37$\pm$0.02 & 0.97$\pm$0.02 & 0.441$\pm$0.014 & 231$\pm$16 & 0.31$\pm$0.02 & 0.98$\pm$0.01 & 0.356$\pm$0.029 \\
\midrule
\quad w/o plasticity term & 44$\pm$1 & 0.20$\pm$0.01 & 0.92$\pm$0.01 & 0.141$\pm$0.008 & 102$\pm$8 & 0.37$\pm$0.01 & 0.98$\pm$0.01 & 0.443$\pm$0.013 & 181$\pm$9 & 0.33$\pm$0.03 & 0.97$\pm$0.01 & 0.361$\pm$0.013 \\
\quad w/o stability term & 59$\pm$1 & 0.20$\pm$0.02 & 0.94$\pm$0.01 & 0.165$\pm$0.011 & 101$\pm$4 & 0.38$\pm$0.01 & 0.97$\pm$0.02 & 0.425$\pm$0.010 & 163$\pm$6 & 0.30$\pm$0.01 & 0.95$\pm$0.05 & 0.290$\pm$0.066 \\
\quad w/o latency term & 154$\pm$127 & 0.17$\pm$0.02 & 0.94$\pm$0.01 & 0.135$\pm$0.021 & 156$\pm$41 & 0.34$\pm$0.01 & 0.99$\pm$0.01 & 0.403$\pm$0.013 & 260$\pm$28 & 0.28$\pm$0.01 & 0.95$\pm$0.02 & 0.267$\pm$0.035 \\
\quad w/o memory term & 66$\pm$5 & 0.20$\pm$0.01 & 0.94$\pm$0.01 & 0.155$\pm$0.005 & 106$\pm$8 & 0.38$\pm$0.01 & 0.97$\pm$0.01 & 0.445$\pm$0.013 & 172$\pm$5 & 0.33$\pm$0.00 & 0.96$\pm$0.03 & 0.335$\pm$0.046 \\
\quad w/o threshold decay & 62$\pm$6 & 0.20$\pm$0.01 & 0.95$\pm$0.01 & 0.170$\pm$0.008 & 111$\pm$1 & 0.37$\pm$0.01 & 0.99$\pm$0.01 & 0.441$\pm$0.023 & 163$\pm$0 & 0.37$\pm$0.00 & 0.99$\pm$0.00 & 0.424$\pm$0.010 \\
\quad w/o shift handling & 57$\pm$3 & 0.19$\pm$0.02 & 0.94$\pm$0.01 & 0.147$\pm$0.013 & 74$\pm$5 & 0.36$\pm$0.01 & 0.98$\pm$0.01 & 0.420$\pm$0.018 & 170$\pm$7 & 0.28$\pm$0.01 & 0.97$\pm$0.01 & 0.295$\pm$0.015 \\
\bottomrule
\end{tabular}}
\end{table*}

\begin{revblock}
\noindent\textbf{Hard memory constraint.} To test strict memory compliance (Sec.~\ref{sec:fine_grained_control}), we run CORe50-NIC, the longest stream and the one on which \Approach{}'s footprint grows most, under a tight budget of 1.5\,GB on the edge server, with and without the projection onto the budget and with and without the memory term of \Healthscore{} (three seeds each). With both safeguards, with only the projection, and with only the memory term, all nine runs complete the 79 experiences within the budget (final accuracy 0.394, 0.401, and 0.415; peak footprint at most 1.45\,GB). Without either safeguard, every run exceeds the budget and is terminated, after 53, 72, and 76 experiences. The projection alone is therefore sufficient for compliance, independently of the value of \Healthscore{}, which is the guarantee the soft term could not give in the submitted version. On this stream the memory term also kept the footprint within budget by itself, but nothing in its form guarantees this in general.

\noindent\textbf{Data prefetching.} On the Jetson AGX Orin (three seeds), we compare three data-loading paths for \Approach{}: no parallel loading, PyTorch DataLoader workers only, and workers plus our prefetcher. Parallel loading matters: without it, the training latency is 297\,s on CORe50-NC and 134\,s on SplitCIFAR100, versus 149\,s and 92\,s with DataLoader workers (a reduction of 50\% and 32\%). Our prefetcher, however, adds no measurable benefit on top of the workers (151\,s and 91\,s); accuracy is unaffected in all three cases. The benefit attributed to prefetching in the submitted version therefore stems from parallel data loading, which DataLoader workers already provide; we keep the prefetcher as an engineering component of the prototype (it gives the stream and the replay buffer a single loading path and is transparent to the algorithms), but no longer count it as a contribution.
\end{revblock}

\begin{revblock}
\subsection{Sensitivity to Hyperparameters}
\label{sec:sensitivity}
Table~\ref{tab:sensitivity} varies each hyperparameter of Table~\ref{tab:hyperparams} by roughly a factor of two around its default while keeping the others fixed. On CORe50-NC, final accuracy stays within $[0.412, 0.449]$ for every setting (default 0.441), i.e., within about three standard deviations of the default. SplitCIFAR100 is more sensitive to the responsiveness of the controller: a smaller batch sensitivity ($\alpha = 1$) or a lower initial threshold ($\Threshold{}_0 = 0.5$) lowers accuracy from 0.165 to 0.127 and 0.126, because the batch size then shrinks too slowly (or grows) during the first experiences, which decide most of the outcome on a ten-experience stream. All other settings stay within $\pm 0.016$ of the default. The defaults are therefore not finely tuned optima, but the controller is robust to moderate changes of every parameter except its two responsiveness parameters on short streams.
\end{revblock}

\begin{table}[!tbp]
\centering\ifrevhighlight\color{blue}\fi
\caption{Sensitivity of \Approach{} to its hyperparameters: final average accuracy (ER, edge server, mean $\pm$ std over three seeds), varying one parameter at a time around the defaults of Table~\ref{tab:hyperparams}.}
\label{tab:sensitivity}
\resizebox{0.85\textwidth}{!}{
\begin{tabular}{llcccccc}
\toprule
 & & \multicolumn{3}{c}{S-CIFAR100 (ACC)} & \multicolumn{3}{c}{CORe50-NC (ACC)} \\
Parameter & Values (low / default / high) & low & default & high & low & default & high \\
\midrule
$\kappa$ & 2 / 4 / 8 & 0.157$\pm$0.010 & 0.165$\pm$0.005 & 0.169$\pm$0.007 & 0.426$\pm$0.013 & 0.441$\pm$0.014 & 0.440$\pm$0.023 \\
$\Threshold{}_0$ & 0.5 / 0.6 / 0.7 & 0.127$\pm$0.003 & 0.165$\pm$0.005 & 0.175$\pm$0.015 & 0.413$\pm$0.019 & 0.441$\pm$0.014 & 0.449$\pm$0.005 \\
$\Threshold{}_{min}$ & 0.3 / 0.4 / 0.5 & 0.149$\pm$0.017 & 0.165$\pm$0.005 & 0.160$\pm$0.008 & 0.430$\pm$0.004 & 0.441$\pm$0.014 & 0.441$\pm$0.010 \\
$\delta$ & 0.1 / 0.3 / 0.6 & 0.158$\pm$0.008 & 0.165$\pm$0.005 & 0.162$\pm$0.004 & 0.445$\pm$0.008 & 0.441$\pm$0.014 & 0.436$\pm$0.013 \\
$\alpha$ & 1 / 2 / 4 & 0.127$\pm$0.008 & 0.165$\pm$0.005 & 0.178$\pm$0.009 & 0.417$\pm$0.014 & 0.441$\pm$0.014 & 0.425$\pm$0.012 \\
$\beta$ & 0.5 / 1 / 2 & 0.162$\pm$0.010 & 0.165$\pm$0.005 & 0.166$\pm$0.003 & 0.433$\pm$0.024 & 0.441$\pm$0.014 & 0.415$\pm$0.028 \\
$h_{safe}$ & 0.1 / 0.15 / 0.25 & 0.157$\pm$0.010 & 0.165$\pm$0.005 & 0.157$\pm$0.005 & 0.436$\pm$0.025 & 0.441$\pm$0.014 & 0.433$\pm$0.016 \\
\bottomrule
\end{tabular}}
\end{table}

\begin{revblock}
\subsection{Other Backbones}
\label{sec:arch}
Table~\ref{tab:arch} repeats the comparison with a deeper residual network (ResNet-56) and a mobile architecture with depthwise-separable convolutions (MobileNetV2). The conclusions carry over. On CORe50-NC, \Approach{} reaches the highest accuracy with both backbones (0.467 and 0.388, versus 0.394 and 0.339 for Miro and 0.340 and 0.255 for MAX-A), with a training latency 12$\times$ lower than MAX-A. On SplitCIFAR100, \Approach{} is best with MobileNetV2 (0.061) and on par with MAX-A with ResNet-56 (0.053 vs.\ 0.058), again an order of magnitude faster. MAX-P's large batches make its footprint grow with model size (10.1\,GB with ResNet-56 and 14.2\,GB with MobileNetV2 on the server), which would not fit an 8\,GB device, whereas \Approach{} stays below 1.6\,GB.
\end{revblock}

\begin{table}[!tbp]
\centering\ifrevhighlight\color{blue}\fi
\caption{Other backbones (ER, edge server, mean $\pm$ std over three seeds): training latency and final average accuracy.}
\label{tab:arch}
\resizebox{\textwidth}{!}{
\begin{tabular}{llcccccccc}
\toprule
 & & \multicolumn{2}{c}{MAX-A} & \multicolumn{2}{c}{MAX-P} & \multicolumn{2}{c}{Miro} & \multicolumn{2}{c}{\Approach{}} \\
Backbone & Benchmark & Lat.\ [s] & ACC & Lat.\ [s] & ACC & Lat.\ [s] & ACC & Lat.\ [s] & ACC \\
\midrule
ResNet-56 & S-CIFAR100 & 1275$\pm$105 & \textbf{0.059$\pm$0.019} & 28$\pm$2 & 0.017$\pm$0.004 & 92$\pm$9 & 0.020$\pm$0.006 & 138$\pm$3 & 0.054$\pm$0.018 \\
ResNet-56 & CORe50-NC & 2519$\pm$116 & 0.340$\pm$0.025 & 53$\pm$1 & 0.205$\pm$0.037 & 216$\pm$20 & 0.395$\pm$0.029 & 212$\pm$24 & \textbf{0.467$\pm$0.008} \\
MobileNetV2 & S-CIFAR100 & 1329$\pm$47 & 0.026$\pm$0.006 & 26$\pm$1 & 0.010$\pm$0.000 & 84$\pm$5 & 0.032$\pm$0.014 & 127$\pm$7 & \textbf{0.062$\pm$0.005} \\
MobileNetV2 & CORe50-NC & 2445$\pm$132 & 0.256$\pm$0.011 & 46$\pm$1 & 0.238$\pm$0.009 & 210$\pm$8 & 0.340$\pm$0.008 & 214$\pm$20 & \textbf{0.389$\pm$0.021} \\
\bottomrule
\end{tabular}}
\end{table}

\subsection{Overhead Analysis}
\label{sec:overhead}

\begin{revblock}
Table~\ref{tab:overhead} reports the cost of \Approach{}'s controller on the Jetson AGX Orin, measured in the same runs as Sec.~\ref{sec:overall_effectiveness}. The controller runs only at experience boundaries; its work consists of computing the indicators, the shift probe (one forward pass over 256 incoming samples), the update of the footprint model, and the reconfiguration of the knobs.

\noindent\textbf{Execution overhead.} The controller takes 1.3--1.8\,s per stream on the four shorter benchmarks, i.e., 1.0--1.5\% of the training time, and 15.1\,s (6.4\%) on CORe50-NIC, whose 79 experiences require 79 decisions and shift probes. The submitted version reported a lower overhead (below 2.1\% on Xavier) because it had neither the shift probe nor the footprint model.

\noindent\textbf{Energy overhead.} Energy is integrated from the board's INA3221 power rails, separately for training and for the controller; the percentages in Table~\ref{tab:overhead} are relative to the training energy of the same run. The controller consumes 0.9--1.2\% of the training energy on the four shorter benchmarks and 6.8\% on CORe50-NIC. This corrects the energy statement of the submitted version, which referred to the wrong quantity (``memory'') and did not state its basis.

\noindent\textbf{Memory overhead.} The controller's own state (set-points, the plugin check's history, and the footprint model) is below 100\,KB. The prefetcher keeps at most two mini-batches in flight on the GPU, i.e., at most $2 \times 2B$ samples, which is 12.6\,MB for $32\times32$ inputs at the maximum batch size $B=256$. Both are included in the measured footprints of all experiments.
\end{revblock}

\begin{table}[!tbp]
\centering\ifrevhighlight\color{blue}\fi
\caption{Overhead of \Approach{}'s controller on Jetson AGX Orin (ER, mean over three seeds): wall-clock time and energy spent at experience boundaries (indicator computation, shift probe, footprint-model update, reconfiguration), relative to the time and energy of training. Energy is integrated from the board's INA3221 power rails.}
\label{tab:overhead}
\resizebox{0.85\textwidth}{!}{
\begin{tabular}{lcccc}
\toprule
Benchmark & Controller time [s] & \% of training time & Controller energy [J] & \% of training energy \\
\midrule
S-CIFAR10 & 1.3 & 1.38\% & 21.6 & 0.93\% \\
S-CIFAR100 & 1.3 & 1.46\% & 22.3 & 1.01\% \\
CORe50-NI & 1.5 & 1.01\% & 42.4 & 0.95\% \\
CORe50-NC & 1.8 & 1.22\% & 49.8 & 1.15\% \\
CORe50-NIC & 15.1 & 6.10\% & 392.5 & 6.51\% \\
\bottomrule
\end{tabular}}
\end{table}

\noindent\begin{minipage}{\linewidth}
\begin{shaded}
    \noindent{\textbf{Low overhead}}: \rev{\Approach{}'s controller costs about 1\% of the training time and energy on typical streams and 6--7\% on a stream with 79 short experiences, and its memory overhead is negligible.}
\end{shaded}
\end{minipage}

\subsection{Discussions}
\label{sec:discussion}

\noindent\textbf{Limitations.}
This work focuses on replay-based OCL under stringent memory constraints, excluding non-replay-based approaches~\citep{kirkpatrick2017overcoming,li2017learning}.  
We do not address OCL for natural language processing tasks, large-scale datasets like ImageNet~\citep{5206848} and \rev{the CLOC benchmark~\citep{cai2021online}}, or infinite streams~\citep{ye2023continual,zhang2024continual}, which remain future work due to resource limitations. 
Following existing continual learning work~\citep{chaudhry2019continual,prabhu2023computationally,lopez2017gradient}, we assume data labels are available for supervised learning, as unsupervised continual learning remains a theoretical challenge. Notably, server-based solutions using large, highly accurate ``golden models''~\citep{bhardwaj2022ekya} could provide annotations, but they are impractical for on-device memory-constrained scenarios due to high memory resource demands.

\rev{\noindent\textbf{Limitations of the controller.} Four limitations emerged from the new experiments. (i) \Approach{} removes offline tuning, not the benefit of it: a static configuration selected from nine offline runs reaches a higher accuracy than \Approach{} on CORe50-NC and CORe50-NIC (Sec.~\ref{sec:ablation}); when such runs are affordable, offline tuning remains the better choice. (ii) The time-dependent threshold, which we kept from the submitted design, does not improve accuracy and hurts it on the longest stream (CORe50-NIC): once the threshold reaches its floor, the controller keeps growing the allocations. A threshold that adapts to the stream length, or no decay at all, is a better default; we did not change it in this revision to keep the evaluation consistent with the design that was reviewed. (iii) The plugin check compares experiences trained with and without the plugins, which differ in content as well; its decisions are therefore noisy on short streams. (iv) The user's importance order has only a small effect on the revised controller (Table~\ref{tab:user_preferences}), because the per-knob indicators already fix the direction of each knob; a controllable trade-off would require mapping preferences to set-points, for example to a latency target.}

\rev{\noindent\textbf{Applicability to vision-language-action models.} VLA models~\citep{brohan2023rt2,kim2024openvla} are an increasingly common policy class for robots, and on-device adaptation of such models is a natural target for \Approach{}. The control problem carries over: adapting a VLA online still trades the memory of activations (batch or sequence length), of replayed episodes, and of optimizer or regularizer state against latency and retention, and the footprint model of Eq.~\ref{eq:footprint} applies unchanged once the per-sample activation cost is calibrated. Three differences matter. First, with billions of parameters, full fine-tuning does not fit an embedded budget at all; on-device adaptation would update a parameter-efficient subset (e.g., LoRA adapters~\citep{hu2021lora}) on top of a frozen, possibly quantized backbone, so the backbone becomes part of $M_{fix}$ and the knobs act on the adapter's working set. Second, replayed samples are multimodal episodes (image sequences, language, and actions), so $M_{df}$ is orders of magnitude larger than for images and the replay knob dominates the budget sooner. Third, plasticity and stability would be measured by task success or action error rather than classification accuracy, which requires a cheap online proxy. We have not evaluated \Approach{} on VLA models; doing so is future work.}

\noindent\textbf{Generality and Broader Impact.} \Approach{} is highly adaptable across diverse datasets, models, and hardware. Built atop \texttt{Avalanche-lib}~\citep{avalanche1,avalanche2}, it seamlessly integrates with scalable DNN training solutions~\citep{jiang2020unified,athlur2022varuna,dean2012large,shoeybi2019megatron,osawa2023pipefisher} for both traditional and resource-constrained environments. Beyond navigational robotics, \Approach{}'s modular and transparent architecture readily extends to other on-device scenarios, including video analytics~\citep{bhardwaj2022ekya}, fault diagnosis~\citep{kim2024design}, user personalization~\citep{chauhan2020contauth,diwan2023continual,pellegrini2021continual}, home automation~\citep{pellegrini2021continual,zhao2022memory}, and smart wearables~\citep{schiemer2023online}. Furthermore, \Approach{} is orthogonal to and fully compatible with existing efficiency techniques. At the model level, it supports automated porting~\citep{guo2021mistify}, model merging~\citep{padmanabhan2023gemel}, compression~\citep{cai2019once,dekhovich2023continual,ren2024esacl,zhang2021impact,geng2021continual,wang2022sparcl,wang2024learn}, and quantization~\citep{karia2024positcl,shi2021continual,ravaglia2021tinyml}. At the system level, it can integrate memory-saving optimizations like \rev{gradient checkpointing~\citep{griewank2000algorithm}}, micro-batch execution~\citep{huang2019gpipe,wang2022melon}, recomputation~\citep{chen2016training,wang2022melon}, and selective layer updates~\citep{wang2019e2,sorrenti2023selective,wang2023egeria}. \rev{Beyond ResNet-20, Sec.~\ref{sec:arch} evaluates a deeper residual network (ResNet-56) and a mobile architecture (MobileNetV2); validating the policies on transformer backbones remains future work.} Ultimately, we hope the engineering insights and design trade-offs will guide improvements in other prominent OCL frameworks~\citep{dimitriadis2023sequel,douillard2021continuum,normandin2021sequoia}, facilitating the practical, widespread deployment of continual learning systems.
\section{Related Work}
\label{sec:related}

Online Continual Learning (OCL) addresses the challenge of sequential learning from streaming data~\citep{chaudhry2019continual,prabhu2023computationally,lopez2017gradient,chaudhry2018efficient,aljundi2019gradient,kirkpatrick2017overcoming,li2017learning,vodisch2023covio,vodisch2023codeps,castri2023continual,soutif2023comprehensive,rebuffi2017icarl}. Replay-based methods, such as ER~\citep{chaudhry2019continual}, mitigate catastrophic forgetting by revisiting past experiences during new task learning\citep{chaudhry2019continual,prabhu2023computationally,chaudhry2018efficient,lopez2017gradient,aljundi2019gradient,aljundi2019online}. These methods perform well across diverse settings and are particularly effective in resource-constrained environments~\citep{prabhu2023computationally}. GEM~\citep{lopez2017gradient} and AGEM~\citep{chaudhry2018efficient} preserve prior knowledge using gradient-based constraints, while GSS~\citep{aljundi2019gradient} enhances memory efficiency by storing informative samples. 
These approaches balance memory retention and adaptability. In practice, OCL serves as an efficient framework for continuous robotic adaptation. It enables autonomous systems to seamlessly adjust to changing lighting and terrains, perform real-time novel object recognition~\citep{pellegrini2020latent}, dynamically update person re-identification models across shifting environments~\citep{ye2024person}, and incrementally perfect physical manipulation tasks~\citep{lee2024incremental}.

Few studies explore system-level optimization for continual learning. For instance, LifeLearner~\citep{kwon2023lifelearner} focuses on hardware-level optimization for meta-continual learning, addressing a distinct problem scope. Ekya~\citep{bhardwaj2022ekya} optimizes offline continual learning, e.g., iCaRL~\citep{rebuffi2017icarl}, for video analytics on high-end multi-GPU servers via dynamic workload scheduling, targeting different systems and applications. Latent Replay~\citep{pellegrini2020latent}, which only modifies algorithms without considering system-level optimization, serves as a strong baseline in our evaluation.

\rev{\noindent\textbf{Closest work: Miro.} Miro~\citep{ma2023cost} is the system closest to \Approach{}. It is a runtime for replay-based on-device continual learning that, at the arrival of each task, profiles candidate configurations online and picks the one with the best accuracy per unit of energy. Concretely, Miro forks the current model, briefly trains each candidate (episodic-memory size and stream-buffer size) on a small slice of the incoming data, discards candidates below an accuracy cutline, and selects the highest accuracy-per-energy configuration; samples are additionally swapped between a small in-memory episodic buffer and a storage-resident memory. \Approach{} differs in four respects. (i) \emph{Mechanism}: Miro is a look-ahead search that spends extra training on throw-away candidate models, whereas \Approach{} is a profiling-free feedback controller that only uses measurements the OCL loop already produces (plasticity, stability, latency, and memory footprint of the previous experience). (ii) \emph{Setting}: Miro targets class-incremental learning with many epochs per task, where a few profiling epochs are a small fraction of training; in single-pass OCL every profiling pass is comparable to the training itself. (iii) \emph{Knobs and constraint}: Miro sizes buffers in samples against an administrator-set sample budget, while \Approach{} co-allocates batch-processing memory, replay memory, and the optimization plugin in bytes and enforces the device memory budget as a hard constraint (Sec.~\ref{sec:fine_grained_control}). (iv) \emph{Objective}: Miro optimizes accuracy per energy explicitly; \Approach{} balances plasticity, stability, and latency under the memory budget and does not optimize energy, which we only report. We re-implement Miro in our pipeline and compare against it directly in Sec.~\ref{sec:miro_comparison}.}

Unlike these prior efforts, which either isolate algorithmic improvements from hardware realities or target resource-rich offline environments, \Approach{} explicitly bridges the gap between OCL algorithms and edge system constraints. \rev{To the best of our knowledge, \Approach{} is the first framework that co-allocates the memory of batch processing, replay, and optimization plugins at runtime to balance training latency and learning efficacy (plasticity and stability) for single-pass online continual learning under a hard device memory budget.} This enables sustainable, real-time robotic adaptation without exceeding the strict memory constraints inherent to low-power SoCs.
\section{Conclusion}

This paper presents \Approach{}, a holistic solution for managing training latency, plasticity, and stability in on-device OCL for memory-limited autonomous embedded systems. Our study highlights the trade-offs required in OCL systems on resource-constrained platforms.  \Approach{} effectively auto-balances these trade-offs while adhering to memory constraints. \rev{In this revision, \Approach{}'s indicator was made dimensionless and evaluated per memory consumer, and memory compliance is guaranteed by a hard constraint. Across five benchmarks, four replay-based algorithms, three backbones, and three memory regimes (an edge server, a Jetson AGX Orin, and the Orin under an 8\,GB budget), with three seeds each, \Approach{} completes every run within the memory budget, is 9--16$\times$ faster than the Avalanche defaults, and reaches the highest final accuracy in most settings, including against Miro, the closest runtime system. A static configuration tuned with several offline runs can still reach a higher accuracy; \Approach{} is the choice when such runs cannot be afforded on the device.} Although we focus on replay-based OCL algorithms, we believe the design principles of  \Approach{} can extend to other continual learning systems, laying a foundation for future on-device OCL systems in intelligent systems. 
\reinbibliography{references}
\clearpage
\appendix
\begin{revblock}
\section{Re-implementation of Miro}
\label{app:miro}

Miro~\citep{ma2023cost} chooses, at the arrival of every task, the size of an in-memory episodic memory and of a stream buffer by online profiling, and swaps samples between the in-memory episodic memory and a storage-resident one. We re-implemented its controller in our Avalanche pipeline; the profiling cost (time and energy) is charged to Miro in all results. The faithful parts are: candidate configurations are sampled uniformly from the feasible grid; every candidate trains a fork of the current model on a 5\% slice of the incoming data; candidates below the top-50\% accuracy cutline are discarded; the selected candidate maximizes accuracy per unit of predicted energy; profiling starts at the second experience. The deviations, all imposed by single-pass online continual learning or by our devices, are:
\begin{enumerate}[leftmargin=14pt]
\item \emph{No storage tier and no swapping.} Our devices keep the whole episodic memory in DRAM; Miro reduces to its configuration profiler.
\item \emph{Stream-buffer knob mapped to the batch size.} Miro does not tune the batch size; in single-pass OCL the stream buffer has no direct analogue, and the batch size is the knob that trades memory for speed.
\item \emph{Single-pass trials from the live model.} Miro trains each candidate for five epochs from a reference checkpoint; in OCL every pass over the incoming data is comparable to the training itself, so each candidate performs one pass over the 5\% slice, starting from the current model, with ten candidates.
\item \emph{Validation data.} The released code evaluates candidates on the test split; we use a held-out slice of buffered and incoming data, which is available on a device.
\item \emph{Energy.} Predicted energy is the trial time per sample, multiplied by the size of the experience and by the average training power measured so far (INA3221 rails on the Jetson, NVML on the server).
\item \emph{No plugin knob.} Miro has no counterpart of the optimization plugins; they stay disabled.
\item \emph{Feasibility.} Candidates are filtered with the same footprint model and budget as \Approach{} instead of a sample-count budget, so both methods face the same memory constraint.
\end{enumerate}
\end{revblock}

\begin{revblock}
\section{Detailed Results}
\label{app:tables}
Tables~\ref{tab:main_server_bench}--\ref{tab:main_orin8_algo} give the numbers behind Figs.~\ref{fig:overall_effectiveness_benchmark} and~\ref{fig:overall_effectiveness_algorithm} (mean $\pm$ standard deviation over seeds 0, 1, and 2). Training latency includes the controller's work at experience boundaries and, for Miro, its profiling.
\end{revblock}

\begin{table*}[!tbp]
\centering\ifrevhighlight\color{blue}\fi
\caption{ER on five benchmarks on the edge server (RTX A4500): mean $\pm$ std over three seeds. OOM: the run exceeded the memory budget (all seeds); T/O: did not finish within 4\,h; a superscript gives the number of seeds that completed. Best value per row in bold.}
\label{tab:main_server_bench}
\resizebox{\textwidth}{!}{
\begin{tabular}{llccccc}
\toprule
Metric & Benchmark & MAX-A & MAX-P & LR & Miro & \Approach{} \\
\midrule
\multirow{5}{*}{Training latency [s] $\downarrow$} & S-CIFAR10 & 720$\pm$114 & \textbf{18$\pm$1} & 618$\pm$90 & 58$\pm$18 & 80$\pm$13 \\
 & S-CIFAR100 & 755$\pm$72 & \textbf{21$\pm$2} & 642$\pm$91 & 70$\pm$20 & 75$\pm$9 \\
 & CORe50-NI & 1406$\pm$99 & \textbf{38$\pm$6} & 1212$\pm$84 & 153$\pm$38 & 120$\pm$15 \\
 & CORe50-NC & 1396$\pm$73 & \textbf{36$\pm$3} & 1212$\pm$81 & 133$\pm$39 & 124$\pm$22 \\
 & CORe50-NIC & 9142$\pm$373 & \textbf{106$\pm$5} & 6620$\pm$262 & 258$\pm$63 & 231$\pm$16 \\
\midrule
\multirow{5}{*}{Plasticity $\uparrow$} & S-CIFAR10 & 0.18$\pm$0.01 & 0.14$\pm$0.04 & 0.17$\pm$0.01 & \textbf{0.31$\pm$0.09} & 0.23$\pm$0.01 \\
 & S-CIFAR100 & 0.11$\pm$0.01 & 0.01$\pm$0.00 & 0.14$\pm$0.03 & 0.08$\pm$0.01 & \textbf{0.19$\pm$0.02} \\
 & CORe50-NI & 0.27$\pm$0.03 & 0.21$\pm$0.02 & 0.24$\pm$0.01 & 0.30$\pm$0.02 & \textbf{0.31$\pm$0.01} \\
 & CORe50-NC & 0.34$\pm$0.02 & 0.24$\pm$0.01 & 0.31$\pm$0.03 & 0.37$\pm$0.01 & \textbf{0.37$\pm$0.02} \\
 & CORe50-NIC & 0.26$\pm$0.01 & 0.19$\pm$0.01 & 0.22$\pm$0.01 & \textbf{0.31$\pm$0.01} & 0.31$\pm$0.02 \\
\midrule
\multirow{5}{*}{Stability $\uparrow$} & S-CIFAR10 & \textbf{0.95$\pm$0.02} & 0.90$\pm$0.04 & 0.93$\pm$0.02 & 0.75$\pm$0.09 & 0.95$\pm$0.03 \\
 & S-CIFAR100 & 0.95$\pm$0.00 & \textbf{1.00$\pm$0.00} & 0.93$\pm$0.03 & 0.96$\pm$0.01 & 0.95$\pm$0.01 \\
 & CORe50-NI & 0.99$\pm$0.01 & 1.00$\pm$0.00 & 1.00$\pm$0.00 & \textbf{1.00$\pm$0.00} & 1.00$\pm$0.00 \\
 & CORe50-NC & 0.96$\pm$0.01 & 0.93$\pm$0.02 & 0.96$\pm$0.02 & 0.96$\pm$0.02 & \textbf{0.97$\pm$0.02} \\
 & CORe50-NIC & 0.95$\pm$0.02 & \textbf{0.98$\pm$0.01} & 0.95$\pm$0.02 & 0.94$\pm$0.02 & 0.98$\pm$0.01 \\
\midrule
\multirow{5}{*}{Final avg.\ accuracy $\uparrow$} & S-CIFAR10 & 0.423$\pm$0.032 & 0.189$\pm$0.024 & 0.286$\pm$0.021 & 0.176$\pm$0.047 & \textbf{0.466$\pm$0.015} \\
 & S-CIFAR100 & 0.104$\pm$0.009 & 0.021$\pm$0.003 & 0.088$\pm$0.004 & 0.067$\pm$0.008 & \textbf{0.165$\pm$0.005} \\
 & CORe50-NI & 0.291$\pm$0.050 & 0.277$\pm$0.041 & 0.264$\pm$0.013 & 0.376$\pm$0.021 & \textbf{0.419$\pm$0.018} \\
 & CORe50-NC & 0.335$\pm$0.028 & 0.214$\pm$0.010 & 0.319$\pm$0.022 & 0.406$\pm$0.039 & \textbf{0.441$\pm$0.014} \\
 & CORe50-NIC & 0.223$\pm$0.032 & 0.250$\pm$0.022 & 0.179$\pm$0.023 & 0.291$\pm$0.032 & \textbf{0.356$\pm$0.029} \\
\bottomrule
\end{tabular}}
\end{table*}

\begin{table*}[!tbp]
\centering\ifrevhighlight\color{blue}\fi
\caption{ER on five benchmarks on Jetson AGX Orin (32\,GB): mean $\pm$ std over three seeds. OOM: the run exceeded the memory budget (all seeds); T/O: did not finish within 4\,h; a superscript gives the number of seeds that completed. Best value per row in bold.}
\label{tab:main_orin_bench}
\resizebox{\textwidth}{!}{
\begin{tabular}{llccccc}
\toprule
Metric & Benchmark & MAX-A & MAX-P & LR & Miro & \Approach{} \\
\midrule
\multirow{5}{*}{Training latency [s] $\downarrow$} & S-CIFAR10 & 746$\pm$11 & \textbf{34$\pm$0} & 568$\pm$2 & 62$\pm$3 & 93$\pm$9 \\
 & S-CIFAR100 & 750$\pm$4 & \textbf{34$\pm$0} & 573$\pm$10 & 57$\pm$2 & 88$\pm$0 \\
 & CORe50-NI & 1434$\pm$4 & \textbf{60$\pm$0} & 1143$\pm$13 & 135$\pm$4 & 154$\pm$31 \\
 & CORe50-NC & 1489$\pm$21 & \textbf{60$\pm$0} & 1181$\pm$15 & 130$\pm$12 & 146$\pm$14 \\
 & CORe50-NIC & T/O & \textbf{158$\pm$1} & 10162$\pm$0 & 202$\pm$0 & 263$\pm$16 \\
\midrule
\multirow{5}{*}{Plasticity $\uparrow$} & S-CIFAR10 & 0.17$\pm$0.01 & 0.15$\pm$0.05 & 0.18$\pm$0.02 & \textbf{0.38$\pm$0.08} & 0.24$\pm$0.02 \\
 & S-CIFAR100 & 0.11$\pm$0.00 & 0.01$\pm$0.00 & 0.13$\pm$0.03 & 0.11$\pm$0.01 & \textbf{0.22$\pm$0.02} \\
 & CORe50-NI & 0.27$\pm$0.03 & 0.21$\pm$0.02 & 0.23$\pm$0.01 & 0.30$\pm$0.02 & \textbf{0.32$\pm$0.02} \\
 & CORe50-NC & 0.34$\pm$0.02 & 0.23$\pm$0.01 & 0.32$\pm$0.01 & \textbf{0.37$\pm$0.01} & 0.36$\pm$0.02 \\
 & CORe50-NIC & T/O & 0.19$\pm$0.00 & 0.20$\pm$0.00 & 0.30$\pm$0.01 & \textbf{0.32$\pm$0.03} \\
\midrule
\multirow{5}{*}{Stability $\uparrow$} & S-CIFAR10 & 0.92$\pm$0.04 & 0.89$\pm$0.04 & 0.92$\pm$0.02 & 0.71$\pm$0.08 & \textbf{0.93$\pm$0.02} \\
 & S-CIFAR100 & 0.96$\pm$0.01 & \textbf{1.00$\pm$0.00} & 0.93$\pm$0.02 & 0.91$\pm$0.02 & 0.93$\pm$0.02 \\
 & CORe50-NI & 0.99$\pm$0.00 & 1.00$\pm$0.00 & 0.99$\pm$0.00 & \textbf{1.00$\pm$0.00} & 1.00$\pm$0.00 \\
 & CORe50-NC & 0.96$\pm$0.02 & 0.93$\pm$0.02 & 0.96$\pm$0.03 & \textbf{0.99$\pm$0.00} & 0.97$\pm$0.01 \\
 & CORe50-NIC & T/O & \textbf{0.98$\pm$0.01} & 0.96$\pm$0.00 & 0.95$\pm$0.01 & 0.97$\pm$0.01 \\
\midrule
\multirow{5}{*}{Final avg.\ accuracy $\uparrow$} & S-CIFAR10 & 0.380$\pm$0.065 & 0.177$\pm$0.017 & 0.296$\pm$0.021 & 0.219$\pm$0.015 & \textbf{0.472$\pm$0.010} \\
 & S-CIFAR100 & 0.091$\pm$0.014 & 0.021$\pm$0.004 & 0.094$\pm$0.006 & 0.044$\pm$0.004 & \textbf{0.162$\pm$0.013} \\
 & CORe50-NI & 0.301$\pm$0.043 & 0.273$\pm$0.046 & 0.249$\pm$0.015 & 0.383$\pm$0.027 & \textbf{0.401$\pm$0.014} \\
 & CORe50-NC & 0.362$\pm$0.033 & 0.220$\pm$0.013 & 0.317$\pm$0.024 & \textbf{0.433$\pm$0.011} & 0.409$\pm$0.023 \\
 & CORe50-NIC & T/O & 0.245$\pm$0.023 & 0.175$\pm$0.000 & 0.294$\pm$0.031 & \textbf{0.323$\pm$0.029} \\
\bottomrule
\end{tabular}}
\end{table*}

\begin{table*}[!tbp]
\centering\ifrevhighlight\color{blue}\fi
\caption{ER on five benchmarks on Jetson AGX Orin under an 8\,GB budget: mean $\pm$ std over three seeds. OOM: the run exceeded the memory budget (all seeds); T/O: did not finish within 4\,h; a superscript gives the number of seeds that completed. Best value per row in bold.}
\label{tab:main_orin8_bench}
\resizebox{\textwidth}{!}{
\begin{tabular}{llccccc}
\toprule
Metric & Benchmark & MAX-A & MAX-P & LR & Miro & \Approach{} \\
\midrule
\multirow{5}{*}{Training latency [s] $\downarrow$} & S-CIFAR10 & 746$\pm$11 & OOM & 568$\pm$2 & \textbf{59$\pm$3} & 94$\pm$9 \\
 & S-CIFAR100 & 750$\pm$4 & OOM & 573$\pm$10 & \textbf{58$\pm$0} & 89$\pm$1 \\
 & CORe50-NI & 1434$\pm$4 & OOM & 1143$\pm$13 & \textbf{133$\pm$8} & 156$\pm$33 \\
 & CORe50-NC & 1489$\pm$21 & OOM & 1181$\pm$15 & \textbf{142$\pm$6} & 158$\pm$10 \\
 & CORe50-NIC & T/O & OOM & OOM & \textbf{205$\pm$2} & 241$\pm$4 \\
\midrule
\multirow{5}{*}{Plasticity $\uparrow$} & S-CIFAR10 & 0.17$\pm$0.01 & OOM & 0.18$\pm$0.02 & \textbf{0.29$\pm$0.03} & 0.26$\pm$0.03 \\
 & S-CIFAR100 & 0.11$\pm$0.00 & OOM & 0.13$\pm$0.03 & 0.10$\pm$0.02 & \textbf{0.19$\pm$0.00} \\
 & CORe50-NI & 0.27$\pm$0.03 & OOM & 0.23$\pm$0.01 & 0.30$\pm$0.02 & \textbf{0.31$\pm$0.01} \\
 & CORe50-NC & 0.34$\pm$0.02 & OOM & 0.32$\pm$0.01 & 0.37$\pm$0.02 & \textbf{0.39$\pm$0.01} \\
 & CORe50-NIC & T/O & OOM & OOM & 0.30$\pm$0.01 & \textbf{0.36$\pm$0.02} \\
\midrule
\multirow{5}{*}{Stability $\uparrow$} & S-CIFAR10 & 0.92$\pm$0.04 & OOM & 0.92$\pm$0.02 & 0.80$\pm$0.08 & \textbf{0.94$\pm$0.02} \\
 & S-CIFAR100 & \textbf{0.96$\pm$0.01} & OOM & 0.93$\pm$0.02 & 0.93$\pm$0.01 & 0.94$\pm$0.00 \\
 & CORe50-NI & 0.99$\pm$0.00 & OOM & 0.99$\pm$0.00 & \textbf{1.00$\pm$0.00} & 0.99$\pm$0.01 \\
 & CORe50-NC & 0.96$\pm$0.02 & OOM & 0.96$\pm$0.03 & \textbf{0.97$\pm$0.00} & 0.94$\pm$0.02 \\
 & CORe50-NIC & T/O & OOM & OOM & 0.97$\pm$0.01 & \textbf{0.97$\pm$0.01} \\
\midrule
\multirow{5}{*}{Final avg.\ accuracy $\uparrow$} & S-CIFAR10 & 0.380$\pm$0.065 & OOM & 0.296$\pm$0.021 & 0.209$\pm$0.042 & \textbf{0.459$\pm$0.025} \\
 & S-CIFAR100 & 0.091$\pm$0.014 & OOM & 0.094$\pm$0.006 & 0.062$\pm$0.005 & \textbf{0.146$\pm$0.013} \\
 & CORe50-NI & 0.301$\pm$0.043 & OOM & 0.249$\pm$0.015 & \textbf{0.408$\pm$0.016} & 0.401$\pm$0.031 \\
 & CORe50-NC & 0.362$\pm$0.033 & OOM & 0.317$\pm$0.024 & 0.419$\pm$0.021 & \textbf{0.431$\pm$0.025} \\
 & CORe50-NIC & T/O & OOM & OOM & 0.331$\pm$0.025 & \textbf{0.385$\pm$0.025} \\
\bottomrule
\end{tabular}}
\end{table*}

\begin{table*}[!tbp]
\centering\ifrevhighlight\color{blue}\fi
\caption{Four OCL algorithms on CORe50-NC on the edge server (RTX A4500): mean $\pm$ std over three seeds. Notation as in Table~\ref{tab:main_server_bench}.}
\label{tab:main_server_algo}
\resizebox{\textwidth}{!}{
\begin{tabular}{llccccc}
\toprule
Metric & Benchmark & MAX-A & MAX-P & LR & Miro & \Approach{} \\
\midrule
\multirow{4}{*}{Training latency [s] $\downarrow$} & ER & 1396$\pm$73 & \textbf{36$\pm$3} & 1212$\pm$81 & 133$\pm$39 & 124$\pm$22 \\
 & GEM & 1991$\pm$191 & \textbf{48$\pm$2} & 1892$\pm$193 & 418$\pm$87 & 324$\pm$42 \\
 & AGEM & 1391$\pm$79 & \textbf{38$\pm$3} & 1109$\pm$127 & 207$\pm$37 & 154$\pm$32 \\
 & GSS & 6066$\pm$560 & \textbf{250$\pm$62} & 4293$\pm$453 & 834$\pm$181 & 747$\pm$56 \\
\midrule
\multirow{4}{*}{Plasticity $\uparrow$} & ER & 0.34$\pm$0.02 & 0.24$\pm$0.01 & 0.31$\pm$0.03 & 0.37$\pm$0.01 & \textbf{0.37$\pm$0.02} \\
 & GEM & 0.32$\pm$0.02 & 0.22$\pm$0.01 & 0.29$\pm$0.02 & \textbf{0.36$\pm$0.02} & 0.34$\pm$0.02 \\
 & AGEM & 0.32$\pm$0.03 & 0.23$\pm$0.01 & 0.30$\pm$0.01 & 0.36$\pm$0.01 & \textbf{0.37$\pm$0.00} \\
 & GSS & 0.20$\pm$0.03 & 0.20$\pm$0.00 & 0.21$\pm$0.01 & 0.30$\pm$0.02 & \textbf{0.33$\pm$0.01} \\
\midrule
\multirow{4}{*}{Stability $\uparrow$} & ER & 0.96$\pm$0.01 & 0.93$\pm$0.02 & 0.96$\pm$0.02 & 0.96$\pm$0.02 & \textbf{0.97$\pm$0.02} \\
 & GEM & \textbf{0.97$\pm$0.00} & 0.96$\pm$0.00 & 0.97$\pm$0.01 & 0.95$\pm$0.04 & 0.96$\pm$0.02 \\
 & AGEM & 0.96$\pm$0.01 & 0.93$\pm$0.01 & \textbf{0.97$\pm$0.00} & 0.95$\pm$0.02 & 0.97$\pm$0.01 \\
 & GSS & 0.92$\pm$0.02 & 0.89$\pm$0.00 & \textbf{0.96$\pm$0.01} & 0.85$\pm$0.02 & 0.94$\pm$0.03 \\
\midrule
\multirow{4}{*}{Final avg.\ accuracy $\uparrow$} & ER & 0.335$\pm$0.028 & 0.214$\pm$0.010 & 0.319$\pm$0.022 & 0.406$\pm$0.039 & \textbf{0.441$\pm$0.014} \\
 & GEM & 0.344$\pm$0.018 & 0.238$\pm$0.017 & 0.317$\pm$0.009 & \textbf{0.375$\pm$0.037} & 0.360$\pm$0.029 \\
 & AGEM & 0.327$\pm$0.034 & 0.228$\pm$0.016 & 0.312$\pm$0.034 & 0.387$\pm$0.019 & \textbf{0.425$\pm$0.015} \\
 & GSS & 0.144$\pm$0.008 & 0.113$\pm$0.012 & 0.182$\pm$0.027 & 0.182$\pm$0.050 & \textbf{0.329$\pm$0.029} \\
\bottomrule
\end{tabular}}
\end{table*}

\begin{table*}[!tbp]
\centering\ifrevhighlight\color{blue}\fi
\caption{Four OCL algorithms on CORe50-NC on Jetson AGX Orin (32\,GB): mean $\pm$ std over three seeds. Notation as in Table~\ref{tab:main_orin_bench}.}
\label{tab:main_orin_algo}
\resizebox{\textwidth}{!}{
\begin{tabular}{llccccc}
\toprule
Metric & Benchmark & MAX-A & MAX-P & LR & Miro & \Approach{} \\
\midrule
\multirow{4}{*}{Training latency [s] $\downarrow$} & ER & 1489$\pm$21 & \textbf{60$\pm$0} & 1181$\pm$15 & 130$\pm$12 & 146$\pm$14 \\
 & GEM & 2076$\pm$21 & \textbf{81$\pm$0} & 1657$\pm$11 & 309$\pm$42 & 328$\pm$82 \\
 & AGEM & 1650$\pm$7 & \textbf{68$\pm$0} & 1323$\pm$7 & 193$\pm$10 & 164$\pm$1 \\
 & GSS & OOM & \textbf{368$\pm$26} & OOM & 1579$\pm$608 & 1025$\pm$27 \\
\midrule
\multirow{4}{*}{Plasticity $\uparrow$} & ER & 0.34$\pm$0.02 & 0.23$\pm$0.01 & 0.32$\pm$0.01 & \textbf{0.37$\pm$0.01} & 0.36$\pm$0.02 \\
 & GEM & 0.34$\pm$0.00 & 0.22$\pm$0.01 & 0.29$\pm$0.01 & 0.33$\pm$0.01 & \textbf{0.35$\pm$0.01} \\
 & AGEM & 0.34$\pm$0.02 & 0.24$\pm$0.00 & 0.31$\pm$0.01 & 0.36$\pm$0.01 & \textbf{0.37$\pm$0.02} \\
 & GSS & OOM & 0.21$\pm$0.01 & OOM & 0.29$\pm$0.02 & \textbf{0.31$\pm$0.01} \\
\midrule
\multirow{4}{*}{Stability $\uparrow$} & ER & 0.96$\pm$0.02 & 0.93$\pm$0.02 & 0.96$\pm$0.03 & \textbf{0.99$\pm$0.00} & 0.97$\pm$0.01 \\
 & GEM & 0.96$\pm$0.02 & 0.96$\pm$0.01 & \textbf{0.98$\pm$0.01} & 0.96$\pm$0.01 & 0.96$\pm$0.01 \\
 & AGEM & 0.96$\pm$0.01 & 0.93$\pm$0.01 & 0.96$\pm$0.02 & 0.97$\pm$0.02 & \textbf{0.97$\pm$0.00} \\
 & GSS & OOM & 0.88$\pm$0.00 & OOM & 0.85$\pm$0.01 & \textbf{0.96$\pm$0.01} \\
\midrule
\multirow{4}{*}{Final avg.\ accuracy $\uparrow$} & ER & 0.362$\pm$0.033 & 0.220$\pm$0.013 & 0.317$\pm$0.024 & \textbf{0.433$\pm$0.011} & 0.409$\pm$0.023 \\
 & GEM & 0.368$\pm$0.011 & 0.229$\pm$0.015 & 0.316$\pm$0.009 & 0.359$\pm$0.027 & \textbf{0.391$\pm$0.022} \\
 & AGEM & 0.356$\pm$0.023 & 0.222$\pm$0.023 & 0.300$\pm$0.035 & \textbf{0.425$\pm$0.019} & 0.417$\pm$0.016 \\
 & GSS & OOM & 0.115$\pm$0.010 & OOM & 0.185$\pm$0.052 & \textbf{0.330$\pm$0.016} \\
\bottomrule
\end{tabular}}
\end{table*}

\begin{table*}[!tbp]
\centering\ifrevhighlight\color{blue}\fi
\caption{Four OCL algorithms on CORe50-NC on Jetson AGX Orin under an 8\,GB budget: mean $\pm$ std over three seeds. Notation as in Table~\ref{tab:main_orin8_bench}.}
\label{tab:main_orin8_algo}
\resizebox{\textwidth}{!}{
\begin{tabular}{llccccc}
\toprule
Metric & Benchmark & MAX-A & MAX-P & LR & Miro & \Approach{} \\
\midrule
\multirow{4}{*}{Training latency [s] $\downarrow$} & ER & 1489$\pm$21 & OOM & 1181$\pm$15 & \textbf{142$\pm$6} & 158$\pm$10 \\
 & GEM & 2076$\pm$21 & OOM & 1657$\pm$11 & \textbf{314$\pm$46} & 334$\pm$84 \\
 & AGEM & 1650$\pm$7 & OOM & 1323$\pm$7 & 191$\pm$10 & \textbf{166$\pm$1} \\
 & GSS & OOM & OOM & OOM & 1656$\pm$5$^{2/3}$ & \textbf{1222$\pm$144} \\
\midrule
\multirow{4}{*}{Plasticity $\uparrow$} & ER & 0.34$\pm$0.02 & OOM & 0.32$\pm$0.01 & 0.37$\pm$0.02 & \textbf{0.39$\pm$0.01} \\
 & GEM & 0.34$\pm$0.00 & OOM & 0.29$\pm$0.01 & 0.34$\pm$0.02 & \textbf{0.38$\pm$0.02} \\
 & AGEM & 0.34$\pm$0.02 & OOM & 0.31$\pm$0.01 & 0.35$\pm$0.01 & \textbf{0.36$\pm$0.01} \\
 & GSS & OOM & OOM & OOM & 0.28$\pm$0.00$^{2/3}$ & \textbf{0.32$\pm$0.01} \\
\midrule
\multirow{4}{*}{Stability $\uparrow$} & ER & 0.96$\pm$0.02 & OOM & 0.96$\pm$0.03 & \textbf{0.97$\pm$0.00} & 0.94$\pm$0.02 \\
 & GEM & 0.96$\pm$0.02 & OOM & \textbf{0.98$\pm$0.01} & 0.94$\pm$0.03 & 0.94$\pm$0.01 \\
 & AGEM & 0.96$\pm$0.01 & OOM & 0.96$\pm$0.02 & 0.96$\pm$0.02 & \textbf{0.98$\pm$0.01} \\
 & GSS & OOM & OOM & OOM & \textbf{0.95$\pm$0.01$^{2/3}$} & 0.92$\pm$0.04 \\
\midrule
\multirow{4}{*}{Final avg.\ accuracy $\uparrow$} & ER & 0.362$\pm$0.033 & OOM & 0.317$\pm$0.024 & 0.419$\pm$0.021 & \textbf{0.431$\pm$0.025} \\
 & GEM & 0.368$\pm$0.011 & OOM & 0.316$\pm$0.009 & 0.351$\pm$0.005 & \textbf{0.409$\pm$0.009} \\
 & AGEM & 0.356$\pm$0.023 & OOM & 0.300$\pm$0.035 & 0.399$\pm$0.023 & \textbf{0.434$\pm$0.026} \\
 & GSS & OOM & OOM & OOM & \textbf{0.303$\pm$0.018$^{2/3}$} & 0.301$\pm$0.044 \\
\bottomrule
\end{tabular}}
\end{table*}


@inproceedings{5206848,
 author = {Deng, Jia and Dong, Wei and Socher, Richard and Li, Li-Jia and Kai Li and Li Fei-Fei},
 booktitle = {2009 IEEE Conference on Computer Vision and Pattern Recognition},
 doi = {10.1109/CVPR.2009.5206848},
 number = {},
 pages = {248-255},
 title = {ImageNet: A large-scale hierarchical image database},
 volume = {},
 year = {2009}
}

@article{aljundi2019gradient,
 author = {Aljundi, Rahaf and Lin, Min and Goujaud, Baptiste and Bengio, Yoshua},
 journal = {Advances in neural information processing systems},
 title = {Gradient based sample selection for online continual learning},
 volume = {32},
 year = {2019}
}

@article{aljundi2019online,
 author = {Aljundi, Rahaf and Belilovsky, Eugene and Tuytelaars, Tinne and Charlin, Laurent and Caccia, Massimo and Lin, Min and Page-Caccia, Lucas},
 journal = {Advances in neural information processing systems},
 title = {Online continual learning with maximal interfered retrieval},
 volume = {32},
 year = {2019}
}

@inproceedings{athlur2022varuna,
 author = {Athlur, Sanjith and Saran, Nitika and Sivathanu, Muthian and Ramjee, Ramachandran and Kwatra, Nipun},
 booktitle = {Proceedings of the Seventeenth European Conference on Computer Systems},
 pages = {472--487},
 title = {Varuna: scalable, low-cost training of massive deep learning models},
 year = {2022}
}

@article{avalanche1,
 author = {Antonio Carta and Lorenzo Pellegrini and Andrea Cossu and Hamed Hemati and Vincenzo Lomonaco},
 journal = {Journal of Machine Learning Research},
 number = {363},
 pages = {1--6},
 title = {Avalanche: A PyTorch Library for Deep Continual Learning},
 url = {http://jmlr.org/papers/v24/23-0130.html},
 volume = {24},
 year = {2023}
}

@inproceedings{avalanche2,
 author = {Vincenzo Lomonaco and Lorenzo Pellegrini and Andrea Cossu and Antonio Carta and Gabriele Graffieti and Tyler L. Hayes and Matthias De Lange and Marc Masana and Jary Pomponi and Gido van de Ven and Martin Mundt and Qi She and Keiland Cooper and Jeremy Forest and Eden Belouadah and Simone Calderara and German I. Parisi and Fabio Cuzzolin and Andreas Tolias and Simone Scardapane and Luca Antiga and Subutai Amhad and Adrian Popescu and Christopher Kanan and Joost van de Weijer and Tinne Tuytelaars and Davide Bacciu and Davide Maltoni},
 booktitle = {Proceedings of IEEE Conference on Computer Vision and Pattern Recognition},
 series = {2nd Continual Learning in Computer Vision Workshop},
 title = {Avalanche: an End-to-End Library for Continual Learning},
 year = {2021}
}

@inproceedings{bhardwaj2022ekya,
 author = {Bhardwaj, Romil and Xia, Zhengxu and Ananthanarayanan, Ganesh and Jiang, Junchen and Shu, Yuanchao and Karianakis, Nikolaos and Hsieh, Kevin and Bahl, Paramvir and Stoica, Ion},
 booktitle = {19th USENIX Symposium on Networked Systems Design and Implementation (NSDI 22)},
 pages = {119--135},
 title = {Ekya: Continuous learning of video analytics models on edge compute servers},
 year = {2022}
}

@misc{bib:orin,
 author = {NVIDIA},
 howpublished = {\url{https://www.nvidia.com/en-us/autonomous-machines/embedded-systems/jetson-orin/}},
 title = {Jetson AGX Orin},
 year = {2022}
}

@inproceedings{brohan2023rt2,
 author = {Zitkovich, Brianna and Yu, Tianhe and Xu, Sichun and Xu, Peng and Xiao, Ted and Xia, Fei and Wu, Jialin and Wohlhart, Paul and Welker, Stefan and Wahid, Ayzaan and others},
 booktitle = {Proceedings of the 7th Conference on Robot Learning (CoRL)},
 pages = {2165--2183},
 series = {Proceedings of Machine Learning Research},
 title = {{RT-2}: Vision-Language-Action Models Transfer Web Knowledge to Robotic Control},
 volume = {229},
 year = {2023}
}

@inproceedings{cai2019once,
 author = {Cai, Han and Gan, Chuang and Wang, Tianzhe and Zhang, Zhekai and Han, Song},
 booktitle = {Proceedings of the International Conference on Learning Representations (ICLR)},
 title = {Once-for-All: Train One Network and Specialize it for Efficient Deployment},
 year = {2020}
}

@inproceedings{cai2021online,
 author = {Cai, Zhipeng and Sener, Ozan and Koltun, Vladlen},
 booktitle = {Proceedings of the IEEE/CVF International Conference on Computer Vision (ICCV)},
 doi = {10.1109/ICCV48922.2021.00817},
 pages = {8261--8270},
 title = {Online Continual Learning with Natural Distribution Shifts: An Empirical Study with Visual Data},
 year = {2021}
}

@inproceedings{castri2023continual,
 author = {Castri, Luca and Mghames, Sariah and Bellotto, Nicola},
 booktitle = {AAAI Bridge Program on Continual Causality},
 organization = {PMLR},
 pages = {85--91},
 title = {From continual learning to causal discovery in robotics},
 year = {2023}
}

@inproceedings{chaudhry2018efficient,
 author = {Chaudhry, Arslan and Ranzato, Marc'Aurelio and Rohrbach, Marcus and Elhoseiny, Mohamed},
 booktitle = {Proceedings of the International Conference on Learning Representations (ICLR)},
 title = {Efficient Lifelong Learning with {A-GEM}},
 year = {2019}
}

@inproceedings{chaudhry2019continual,
 author = {Chaudhry, Arslan and Rohrbach, Marcus and Elhoseiny, Mohamed and Ajanthan, Thalaiyasingam and Dokania, P and Torr, P and Ranzato, M},
 booktitle = {Workshop on Multi-Task and Lifelong Reinforcement Learning},
 title = {Continual learning with tiny episodic memories},
 year = {2019}
}

@article{chauhan2020contauth,
 author = {Chauhan, Jagmohan and Kwon, Young D and Hui, Pan and Mascolo, Cecilia},
 journal = {Proceedings of the ACM on Interactive, Mobile, Wearable and Ubiquitous Technologies},
 number = {4},
 pages = {1--23},
 publisher = {ACM New York, NY, USA},
 title = {Contauth: Continual learning framework for behavioral-based user authentication},
 volume = {4},
 year = {2020}
}

@misc{chen2016training,
 author = {Chen, Tianqi and Xu, Bing and Zhang, Chiyuan and Guestrin, Carlos},
 howpublished = {arXiv preprint arXiv:1604.06174},
 title = {Training Deep Nets with Sublinear Memory Cost},
 year = {2016}
}

@inproceedings{dean2012large,
 author = {Dean, Jeffrey and Corrado, Greg and Monga, Rajat and Chen, Kai and Devin, Matthieu and Mao, Mark and Ranzato, Marc'aurelio and Senior, Andrew and Tucker, Paul and Yang, Ke and others},
 booktitle = {Advances in Neural Information Processing Systems (NeurIPS)},
 title = {Large Scale Distributed Deep Networks},
 volume = {25},
 year = {2012}
}

@article{dekhovich2023continual,
 author = {Dekhovich, Aleksandr and Tax, David MJ and Sluiter, Marcel HF and Bessa, Miguel A},
 journal = {Applied Intelligence},
 number = {14},
 pages = {17849--17864},
 publisher = {Springer},
 title = {Continual prune-and-select: class-incremental learning with specialized subnetworks},
 volume = {53},
 year = {2023}
}

@misc{dimitriadis2023sequel,
 author = {Dimitriadis, Nikolaos and Fleuret, Fran{\c{c}}ois and Frossard, Pascal},
 howpublished = {arXiv preprint arXiv:2304.10857},
 title = {{SequeL}: A Continual Learning Library in {PyTorch} and {JAX}},
 year = {2023}
}

@inproceedings{diwan2023continual,
 author = {Diwan, Anuj and Yeh, Ching-Feng and Hsu, Wei-Ning and Tomasello, Paden and Choi, Eunsol and Harwath, David and Mohamed, Abdelrahman},
 booktitle = {ICASSP 2023-2023 IEEE International Conference on Acoustics, Speech and Signal Processing (ICASSP)},
 organization = {IEEE},
 pages = {1--5},
 title = {Continual learning for on-device speech recognition using disentangled conformers},
 year = {2023}
}

@misc{douillard2021continuum,
 author = {Douillard, Arthur and Lesort, Timoth{\'e}e},
 howpublished = {arXiv preprint arXiv:2102.06253},
 title = {Continuum: Simple Management of Complex Continual Learning Scenarios},
 year = {2021}
}

@inproceedings{geng2021continual,
 author = {Geng, Binzong and Yuan, Fajie and Xu, Qiancheng and Shen, Ying and Xu, Ruifeng and Yang, Min},
 booktitle = {Proceedings of the 59th Annual Meeting of the Association for Computational Linguistics and the 11th International Joint Conference on Natural Language Processing (ACL-IJCNLP), Volume 2: Short Papers},
 doi = {10.18653/v1/2021.acl-short.66},
 pages = {517--523},
 title = {Continual Learning for Task-oriented Dialogue System with Iterative Network Pruning, Expanding and Masking},
 year = {2021}
}

@inproceedings{ghunaim2023real,
 author = {Ghunaim, Yasir and Bibi, Adel and Alhamoud, Kumail and Alfarra, Motasem and Al Kader Hammoud, Hasan Abed and Prabhu, Ameya and Torr, Philip HS and Ghanem, Bernard},
 booktitle = {Proceedings of the IEEE/CVF Conference on Computer Vision and Pattern Recognition},
 pages = {11888--11897},
 title = {Real-time evaluation in online continual learning: A new hope},
 year = {2023}
}

@misc{goyal2017accurate,
 author = {Goyal, Priya and Doll{\'a}r, Piotr and Girshick, Ross and Noordhuis, Pieter and Wesolowski, Lukasz and Kyrola, Aapo and Tulloch, Andrew and Jia, Yangqing and He, Kaiming},
 howpublished = {arXiv preprint arXiv:1706.02677},
 title = {Accurate, Large Minibatch {SGD}: Training {ImageNet} in 1 Hour},
 year = {2017}
}

@article{griewank2000algorithm,
 author = {Griewank, Andreas and Walther, Andrea},
 journal = {ACM Transactions on Mathematical Software (TOMS)},
 number = {1},
 pages = {19--45},
 publisher = {ACM New York, NY, USA},
 title = {Algorithm 799: revolve: an implementation of checkpointing for the reverse or adjoint mode of computational differentiation},
 volume = {26},
 year = {2000}
}

@inproceedings{guo2021mistify,
 author = {Guo, Peizhen and Hu, Bo and Hu, Wenjun},
 booktitle = {18th USENIX Symposium on Networked Systems Design and Implementation (NSDI 21)},
 pages = {705--719},
 title = {Mistify: Automating $\{$DNN$\}$ Model Porting for $\{$On-Device$\}$ Inference at the Edge},
 year = {2021}
}

@inproceedings{hajizada2024continual,
 author = {Hajizada, Elvin and Swaminathan, Balachandran and Sandamirskaya, Yulia},
 booktitle = {Proceedings of the IEEE/RSJ International Conference on Intelligent Robots and Systems (IROS)},
 doi = {10.1109/IROS58592.2024.10802683},
 pages = {13988--13995},
 title = {Continual Learning for Autonomous Robots: A Prototype-based Approach},
 year = {2024}
}

@inproceedings{he2016deep,
 author = {He, Kaiming and Zhang, Xiangyu and Ren, Shaoqing and Sun, Jian},
 booktitle = {Proceedings of the IEEE conference on computer vision and pattern recognition},
 pages = {770--778},
 title = {Deep residual learning for image recognition},
 year = {2016}
}

@inproceedings{hess2021procedural,
 author = {Hess, Timm and Mundt, Martin and Pliushch, Iuliia and Ramesh, Visvanathan},
 booktitle = {Proceedings of the Neural Information Processing Systems Track on Datasets and Benchmarks (NeurIPS Datasets and Benchmarks)},
 title = {A Procedural World Generation Framework for Systematic Evaluation of Continual Learning},
 volume = {1},
 year = {2021}
}

@inproceedings{hu2021lora,
 author = {Hu, Edward J. and Shen, Yelong and Wallis, Phillip and Allen-Zhu, Zeyuan and Li, Yuanzhi and Wang, Shean and Wang, Lu and Chen, Weizhu},
 booktitle = {Proceedings of the International Conference on Learning Representations (ICLR)},
 title = {{LoRA}: Low-Rank Adaptation of Large Language Models},
 year = {2022}
}

@article{huang2019gpipe,
 author = {Huang, Yanping and Cheng, Youlong and Bapna, Ankur and Firat, Orhan and Chen, Dehao and Chen, Mia and Lee, HyoukJoong and Ngiam, Jiquan and Le, Quoc V and Wu, Yonghui and others},
 journal = {Advances in neural information processing systems},
 title = {Gpipe: Efficient training of giant neural networks using pipeline parallelism},
 volume = {32},
 year = {2019}
}

@inproceedings{jiang2020unified,
 author = {Jiang, Yimin and Zhu, Yibo and Lan, Chang and Yi, Bairen and Cui, Yong and Guo, Chuanxiong},
 booktitle = {14th USENIX Symposium on Operating Systems Design and Implementation (OSDI 20)},
 pages = {463--479},
 title = {A unified architecture for accelerating distributed $\{$DNN$\}$ training in heterogeneous $\{$GPU/CPU$\}$ clusters},
 year = {2020}
}

@inproceedings{karia2024positcl,
 author = {Karia, Vedant and Zyarah, Abdullah and Kudithipudi, Dhireesha},
 booktitle = {Proceedings of the Great Lakes Symposium on VLSI 2024},
 pages = {645--650},
 title = {PositCL: Compact Continual Learning with Posit Aware Quantization},
 year = {2024}
}

@inproceedings{kato2018autoware,
 author = {Kato, Shinpei and Tokunaga, Shota and Maruyama, Yuya and Maeda, Seiya and Hirabayashi, Manato and Kitsukawa, Yuki and Monrroy, Abraham and Ando, Tomohito and Fujii, Yusuke and Azumi, Takuya},
 booktitle = {2018 ACM/IEEE 9th International Conference on Cyber-Physical Systems (ICCPS)},
 organization = {IEEE},
 pages = {287--296},
 title = {Autoware on board: Enabling autonomous vehicles with embedded systems},
 year = {2018}
}

@article{kim2024design,
 author = {Kim, Youngjun and Kim, Taewan and Kim, Suhyun and Lee, Seongjae and Kim, Taehyoun},
 journal = {IEMEK Journal of Embedded Systems and Applications},
 number = {3},
 pages = {151--158},
 publisher = {Institute of Embedded Engineering of Korea},
 title = {Design and Implementation of a Lightweight On-Device AI-Based Real-time Fault Diagnosis System using Continual Learning},
 volume = {19},
 year = {2024}
}

@inproceedings{kim2024openvla,
 author = {Kim, Moo Jin and Pertsch, Karl and Karamcheti, Siddharth and Xiao, Ted and Balakrishna, Ashwin and Nair, Suraj and Rafailov, Rafael and Foster, Ethan and Lam, Grace and Sanketi, Pannag and Vuong, Quan and Kollar, Thomas and Burchfiel, Benjamin and Tedrake, Russ and Sadigh, Dorsa and Levine, Sergey and Liang, Percy and Finn, Chelsea},
 booktitle = {Proceedings of the 8th Conference on Robot Learning (CoRL)},
 pages = {2679--2713},
 series = {Proceedings of Machine Learning Research},
 title = {{OpenVLA}: An Open-Source Vision-Language-Action Model},
 volume = {270},
 year = {2024}
}

@article{kirkpatrick2017overcoming,
 author = {Kirkpatrick, James and Pascanu, Razvan and Rabinowitz, Neil and Veness, Joel and Desjardins, Guillaume and Rusu, Andrei A and Milan, Kieran and Quan, John and Ramalho, Tiago and Grabska-Barwinska, Agnieszka and others},
 journal = {Proceedings of the national academy of sciences},
 number = {13},
 pages = {3521--3526},
 publisher = {National Acad Sciences},
 title = {Overcoming catastrophic forgetting in neural networks},
 volume = {114},
 year = {2017}
}

@inproceedings{kisavcanin2017deep,
 author = {Kisa{\v{c}}anin, Branislav},
 booktitle = {2017 IEEE 47th International Symposium on Multiple-Valued Logic (ISMVL)},
 organization = {IEEE},
 pages = {142--142},
 title = {Deep learning for autonomous vehicles},
 year = {2017}
}

@techreport{krizhevsky2009learning,
 address = {Toronto, ON, Canada},
 author = {Krizhevsky, Alex and Hinton, Geoffrey},
 institution = {University of Toronto},
 title = {Learning Multiple Layers of Features from Tiny Images},
 year = {2009}
}

@inproceedings{kwon2023lifelearner,
 author = {Kwon, Young D. and Chauhan, Jagmohan and Jia, Hong and Venieris, Stylianos I. and Mascolo, Cecilia},
 booktitle = {Proceedings of the 21st ACM Conference on Embedded Networked Sensor Systems (SenSys)},
 doi = {10.1145/3625687.3625804},
 pages = {138--151},
 title = {{LifeLearner}: Hardware-Aware Meta Continual Learning System for Embedded Computing Platforms},
 year = {2023}
}

@article{lee2024incremental,
 author = {Lee, Daehee and Yoo, Minjong and Kim, Woo Kyung and Choi, Wonje and Woo, Honguk},
 journal = {Advances in Neural Information Processing Systems},
 pages = {17286--17312},
 title = {Incremental learning of retrievable skills for efficient continual task adaptation},
 volume = {37},
 year = {2024}
}

@article{li2017learning,
 author = {Li, Zhizhong and Hoiem, Derek},
 journal = {IEEE transactions on pattern analysis and machine intelligence},
 number = {12},
 pages = {2935--2947},
 publisher = {IEEE},
 title = {Learning without forgetting},
 volume = {40},
 year = {2017}
}

@inproceedings{lomanco2017core50,
 author = {Lomonaco, Vincenzo and Maltoni, Davide},
 booktitle = {Proceedings of the 1st Annual Conference on Robot Learning (CoRL)},
 pages = {17--26},
 series = {Proceedings of Machine Learning Research},
 title = {{CORe50}: a New Dataset and Benchmark for Continuous Object Recognition},
 volume = {78},
 year = {2017}
}

@inproceedings{lomonaco2019fine,
 author = {Lomonaco, Vincenzo and Maltoni, Davide and Pellegrini, Lorenzo},
 booktitle = {Proceedings of the IEEE/CVF Conference on Computer Vision and Pattern Recognition Workshops (CVPRW)},
 doi = {10.1109/CVPRW50498.2020.00131},
 pages = {989--998},
 title = {Rehearsal-Free Continual Learning over Small Non-{I.I.D.} Batches},
 year = {2020}
}

@article{lopez2017gradient,
 author = {Lopez-Paz, David and Ranzato, Marc'Aurelio},
 journal = {Advances in neural information processing systems},
 title = {Gradient episodic memory for continual learning},
 volume = {30},
 year = {2017}
}

@inproceedings{ma2023cost,
 author = {Ma, Xinyue and Jeong, Suyeon and Zhang, Minjia and Wang, Di and Choi, Jonghyun and Jeon, Myeongjae},
 booktitle = {Proceedings of the 29th Annual International Conference on Mobile Computing and Networking (MobiCom)},
 doi = {10.1145/3570361.3613297},
 pages = {1--15},
 title = {Cost-effective On-device Continual Learning over Memory Hierarchy with {Miro}},
 year = {2023}
}

@article{mai2022online,
 author = {Mai, Zheda and Li, Ruiwen and Jeong, Jihwan and Quispe, David and Kim, Hyunwoo and Sanner, Scott},
 journal = {Neurocomputing},
 pages = {28--51},
 publisher = {Elsevier},
 title = {Online continual learning in image classification: An empirical survey},
 volume = {469},
 year = {2022}
}

@article{nie2023online,
 author = {Nie, Xiangli and Deng, Zhiguang and He, Mingdong and Fan, Mingyu and Tang, Zheng},
 journal = {IEEE Transactions on Neural Networks and Learning Systems},
 publisher = {IEEE},
 title = {Online active continual learning for robotic lifelong object recognition},
 year = {2023}
}

@misc{normandin2021sequoia,
 author = {Normandin, Fabrice and Golemo, Florian and Ostapenko, Oleksiy and Rodriguez, Pau and Riemer, Matthew D and Hurtado, Julio and Khetarpal, Khimya and Lindeborg, Ryan and Cecchi, Lucas and Lesort, Timoth{\'e}e and others},
 howpublished = {arXiv preprint arXiv:2108.01005},
 title = {Sequoia: A Software Framework to Unify Continual Learning Research},
 year = {2021}
}

@book{odum1971fundamentals,
 author = {Odum, Eugene Pleasants and Barrett, Gary W and others},
 publisher = {Saunders Philadelphia},
 title = {Fundamentals of ecology},
 volume = {3},
 year = {1971}
}

@article{osawa2023pipefisher,
 author = {Osawa, Kazuki and Li, Shigang and Hoefler, Torsten},
 journal = {Proceedings of Machine Learning and Systems},
 title = {Pipefisher: Efficient training of large language models using pipelining and fisher information matrices},
 volume = {5},
 year = {2023}
}

@inproceedings{padmanabhan2023gemel,
 author = {Padmanabhan, Arthi and Agarwal, Neil and Iyer, Anand and Ananthanarayanan, Ganesh and Shu, Yuanchao and Karianakis, Nikolaos and Xu, Guoqing Harry and Netravali, Ravi},
 booktitle = {20th USENIX Symposium on Networked Systems Design and Implementation (NSDI 23)},
 pages = {973--994},
 title = {Gemel: Model Merging for $\{$Memory-Efficient$\}$,$\{$Real-Time$\}$ Video Analytics at the Edge},
 year = {2023}
}

@inproceedings{pellegrini2020latent,
 author = {Pellegrini, Lorenzo and Graffieti, Gabriele and Lomonaco, Vincenzo and Maltoni, Davide},
 booktitle = {2020 IEEE/RSJ International Conference on Intelligent Robots and Systems (IROS)},
 organization = {IEEE},
 pages = {10203--10209},
 title = {Latent replay for real-time continual learning},
 year = {2020}
}

@inproceedings{pellegrini2021continual,
 author = {Pellegrini, Lorenzo and Lomonaco, Vincenzo and Graffieti, Gabriele and Maltoni, Davide},
 booktitle = {Proceedings of the European Symposium on Artificial Neural Networks, Computational Intelligence and Machine Learning (ESANN)},
 doi = {10.14428/esann/2021.es2021-136},
 pages = {23--28},
 title = {Continual Learning at the Edge: Real-Time Training on Smartphone Devices},
 year = {2021}
}

@inproceedings{popov2022nvradarnet,
 author = {Popov, Alexander and Gebhardt, Patrik and Chen, Ke and Oldja, Ryan and Lee, Heeseok and Murray, Shane and Bhargava, Ruchi and Smolyanskiy, Nikolai},
 booktitle = {Proceedings of the IEEE International Conference on Robotics and Automation (ICRA)},
 doi = {10.1109/ICRA48891.2023.10160592},
 pages = {6958--6964},
 title = {{NVRadarNet}: Real-Time Radar Obstacle and Free Space Detection for Autonomous Driving},
 year = {2023}
}

@inproceedings{prabhu2023computationally,
 author = {Prabhu, Ameya and Al Kader Hammoud, Hasan Abed and Dokania, Puneet K and Torr, Philip HS and Lim, Ser-Nam and Ghanem, Bernard and Bibi, Adel},
 booktitle = {Proceedings of the IEEE/CVF Conference on Computer Vision and Pattern Recognition},
 pages = {3698--3707},
 title = {Computationally budgeted continual learning: What does matter?},
 year = {2023}
}

@article{ravaglia2021tinyml,
 author = {Ravaglia, Leonardo and Rusci, Manuele and Nadalini, Davide and Capotondi, Alessandro and Conti, Francesco and Benini, Luca},
 journal = {IEEE Journal on Emerging and Selected Topics in Circuits and Systems},
 number = {4},
 pages = {789--802},
 publisher = {IEEE},
 title = {A tinyml platform for on-device continual learning with quantized latent replays},
 volume = {11},
 year = {2021}
}

@inproceedings{rebuffi2017icarl,
 author = {Rebuffi, Sylvestre-Alvise and Kolesnikov, Alexander and Sperl, Georg and Lampert, Christoph H},
 booktitle = {Proceedings of the IEEE conference on Computer Vision and Pattern Recognition},
 pages = {2001--2010},
 title = {icarl: Incremental classifier and representation learning},
 year = {2017}
}

@inproceedings{ren2024esacl,
 author = {Ren, Weijieying and Honavar, Vasant G.},
 booktitle = {Proceedings of the SIAM International Conference on Data Mining (SDM)},
 doi = {10.1137/1.9781611978032.19},
 pages = {163--171},
 title = {{EsaCL}: An Efficient Continual Learning Algorithm},
 year = {2024}
}

@manual{robotisemanual2024,
 author = {{ROBOTIS}},
 note = {Accessed: 15 April 2024},
 organization = {ROBOTIS},
 title = {TurtleBot3 Overview},
 url = {https://emanual.robotis.com/docs/en/platform/turtlebot3/overview/},
 year = {2024}
}

@article{schiemer2023online,
 author = {Schiemer, Martin and Fang, Lei and Dobson, Simon and Ye, Juan},
 journal = {Pervasive and Mobile Computing},
 pages = {101817},
 publisher = {Elsevier},
 title = {Online continual learning for human activity recognition},
 volume = {93},
 year = {2023}
}

@inproceedings{shi2021continual,
 author = {Shi, Yujun and Yuan, Li and Chen, Yunpeng and Feng, Jiashi},
 booktitle = {Proceedings of the IEEE/CVF conference on Computer Vision and Pattern Recognition},
 pages = {16674--16683},
 title = {Continual learning via bit-level information preserving},
 year = {2021}
}

@misc{shoeybi2019megatron,
 author = {Shoeybi, Mohammad and Patwary, Mostofa and Puri, Raul and LeGresley, Patrick and Casper, Jared and Catanzaro, Bryan},
 howpublished = {arXiv preprint arXiv:1909.08053},
 title = {{Megatron-LM}: Training Multi-Billion Parameter Language Models Using Model Parallelism},
 year = {2019}
}

@inproceedings{sorrenti2023selective,
 author = {Sorrenti, Amelia and Bellitto, Giovanni and Salanitri, Federica Proietto and Pennisi, Matteo and Spampinato, Concetto and Palazzo, Simone},
 booktitle = {Proceedings of the IEEE/CVF International Conference on Computer Vision},
 pages = {3550--3559},
 title = {Selective Freezing for Efficient Continual Learning},
 year = {2023}
}

@inproceedings{soutif2023comprehensive,
 author = {Soutif-Cormerais, Albin and Carta, Antonio and Cossu, Andrea and Hurtado, Julio and Lomonaco, Vincenzo and Van de Weijer, Joost and Hemati, Hamed},
 booktitle = {Proceedings of the IEEE/CVF International Conference on Computer Vision},
 pages = {3518--3528},
 title = {A comprehensive empirical evaluation on online continual learning},
 year = {2023}
}

@misc{SparkFun_JetBot,
 author = {{SparkFun Electronics}},
 howpublished = {\url{https://www.sparkfun.com/products/18486}},
 title = {SparkFun JetBot AI Kit},
 year = {2022}
}

@inproceedings{vodisch2023codeps,
 author = {V{\"o}disch, Niclas and Petek, K{\"u}rsat and Burgard, Wolfram and Valada, Abhinav},
 booktitle = {Proceedings of Robotics: Science and Systems (RSS)},
 doi = {10.15607/RSS.2023.XIX.073},
 title = {{CoDEPS}: Online Continual Learning for Depth Estimation and Panoptic Segmentation},
 volume = {XIX},
 year = {2023}
}

@inproceedings{vodisch2023covio,
 author = {V{\"o}disch, Niclas and Cattaneo, Daniele and Burgard, Wolfram and Valada, Abhinav},
 booktitle = {Proceedings of the IEEE/CVF Conference on Computer Vision and Pattern Recognition},
 pages = {2464--2473},
 title = {Covio: Online continual learning for visual-inertial odometry},
 year = {2023}
}

@article{wang2019e2,
 author = {Wang, Yue and Jiang, Ziyu and Chen, Xiaohan and Xu, Pengfei and Zhao, Yang and Lin, Yingyan and Wang, Zhangyang},
 journal = {Advances in Neural Information Processing Systems},
 title = {E2-train: Training state-of-the-art cnns with over 80\% energy savings},
 volume = {32},
 year = {2019}
}

@inproceedings{wang2022melon,
 author = {Wang, Qipeng and Xu, Mengwei and Jin, Chao and Dong, Xinran and Yuan, Jinliang and Jin, Xin and Huang, Gang and Liu, Yunxin and Liu, Xuanzhe},
 booktitle = {Proceedings of the 20th Annual International Conference on Mobile Systems, Applications and Services},
 pages = {450--463},
 title = {Melon: Breaking the memory wall for resource-efficient on-device machine learning},
 year = {2022}
}

@article{wang2022sparcl,
 author = {Wang, Zifeng and Zhan, Zheng and Gong, Yifan and Yuan, Geng and Niu, Wei and Jian, Tong and Ren, Bin and Ioannidis, Stratis and Wang, Yanzhi and Dy, Jennifer},
 journal = {Advances in Neural Information Processing Systems},
 pages = {20366--20380},
 title = {Sparcl: Sparse continual learning on the edge},
 volume = {35},
 year = {2022}
}

@inproceedings{wang2023egeria,
 author = {Wang, Yiding and Sun, Decang and Chen, Kai and Lai, Fan and Chowdhury, Mosharaf},
 booktitle = {Proceedings of the Eighteenth European Conference on Computer Systems},
 pages = {851--866},
 title = {Egeria: Efficient dnn training with knowledge-guided layer freezing},
 year = {2023}
}

@inproceedings{wang2024learn,
 author = {Wang, Mingyang and Adel, Heike and Lange, Lukas and Str{\"o}tgen, Jannik and Sch{\"u}tze, Hinrich},
 booktitle = {Proceedings of the 9th Workshop on Representation Learning for NLP (RepL4NLP)},
 doi = {10.18653/v1/2024.repl4nlp-1.12},
 pages = {163--176},
 title = {Learn it or Leave it: Module Composition and Pruning for Continual Learning},
 year = {2024}
}

@misc{Waveshare_JetBot,
 author = {{Waveshare}},
 howpublished = {\url{https://www.amazon.com/Waveshare-JetBot-AI-Kit-Accessories/dp/B07V8JL4TF/}},
 title = {Waveshare JetBot AI Kit},
 year = {2022}
}

@inproceedings{ye2023continual,
 author = {Ye, Fei and Bors, Adrian G},
 booktitle = {Proceedings of the AAAI Conference on Artificial Intelligence},
 number = {9},
 pages = {10918--10926},
 title = {Continual variational autoencoder via continual generative knowledge distillation},
 volume = {37},
 year = {2023}
}

@article{ye2024person,
 author = {Ye, Hanjing and Zhao, Jieting and Zhan, Yu and Chen, Weinan and He, Li and Zhang, Hong},
 journal = {IEEE Robotics and Automation Letters},
 publisher = {IEEE},
 title = {Person re-identification for robot person following with online continual learning},
 year = {2024}
}

@inproceedings{zhang2021impact,
 author = {Zhang, Xueyang and Li, Hang and Chen, Xi and Liu, Xue},
 booktitle = {2021 IEEE Third International Conference on Cognitive Machine Intelligence (CogMI)},
 organization = {IEEE},
 pages = {27--33},
 title = {Impact Patterns of Combining Model Pruning and Continual Learning on Model Performance},
 year = {2021}
}

@inproceedings{zhang2024continual,
 author = {Zhang, Wenxuan and Mohamed, Youssef and Ghanem, Bernard and Torr, Philip H. S. and Bibi, Adel and Elhoseiny, Mohamed},
 booktitle = {Proceedings of the International Conference on Learning Representations (ICLR)},
 title = {Continual Learning on a Diet: Learning from Sparsely Labeled Streams Under Constrained Computation},
 year = {2024}
}

@inproceedings{zhao2022memory,
 author = {Zhao, Yuqing and Saxena, Divya and Cao, Jiannong},
 booktitle = {Proceedings of the 20th ACM Conference on Embedded Networked Sensor Systems},
 pages = {1175--1181},
 title = {Memory-efficient domain incremental learning for Internet of Things},
 year = {2022}
}
\end{document}